\documentclass[aps,pra,reprint,nofootinbib,longbibliography,floatfix]{revtex4-2}
\usepackage[T1]{fontenc}
\usepackage[utf8]{inputenc}
\usepackage{lmodern}
\usepackage{amsmath,amssymb,amsthm,mathtools,bm}
\usepackage{microtype,graphicx}
\usepackage[dvipsnames]{xcolor}
\usepackage{tikz}
\usetikzlibrary{arrows.meta}
\usepackage[colorlinks=true,
            linkcolor=MidnightBlue,
            citecolor=MidnightBlue,
            urlcolor=MidnightBlue]{hyperref}

\graphicspath{{figures/}}

\newtheorem{theorem}{Theorem}
\newtheorem{corollary}[theorem]{Corollary}
\newtheorem{proposition}[theorem]{Proposition}
\newtheorem{lemma}[theorem]{Lemma}

\DeclareMathOperator{\Tr}{Tr}
\DeclareMathOperator{\SEP}{SEP}

\DeclareMathOperator{\Bin}{Bin}

\newcommand{\id}{\mathrm{id}}
\newcommand{\ER}{E_R}
\newcommand{\ERi}{E_R^{\infty}}
\newcommand{\ERinf}{E_R^{\infty}}

\newcommand{\HH}{\mathcal H}
\newcommand{\PP}{\mathcal P}
\newcommand{\ZZ}{\mathbb Z}
\newcommand{\EE}{\mathbb E}

\newcommand{\ket}[1]{\lvert #1\rangle}
\newcommand{\bra}[1]{\langle #1\rvert}
\newcommand{\proj}[1]{\ket{#1}\!\bra{#1}}

\newcommand{\AD}{\mathcal A_{\gamma}}
\newcommand{\Bsec}{B_{\mathrm{sec}}}
\newcommand{\UPLOB}{U_{\mathrm{PLOB}}}
\newcommand{\Umax}{U_{\max}}
\newcommand{\Usq}{U_{\mathrm{sq}}}
\newcommand{\RR}{R_{\mathrm{RCI}}}
\newcommand{\GAD}{\mathcal G_{\gamma,q}}
\newcommand{\Umix}{U_{\mathrm{mix}}}

\begin{document}

\title{Improved two-way capacity upper bounds \\ for amplitude-damping channels}
\author{Stefano Pirandola}
\affiliation{Department of Computer Science, University of York, York YO10 5GH, United Kingdom}

\begin{abstract}
We derive an improved upper bound on the two-way-assisted capacities of the qubit amplitude-damping channel with decay probability $\gamma$, given by $\log_2[1+(1-\gamma)\gamma^{\gamma/(1-\gamma)}]$. It applies to quantum communication, entanglement distribution, private communication, and secret-key agreement, and, for every $0<\gamma<1$, strictly improves the PLOB relative-entropy bound, the analytic max-relative-entropy bound, and the optimized balanced squashed-entanglement bound. At high damping, a known triple-rail protocol attains more than $99.54\%$ of its leading coefficient. Our method extends sector teleportation simulation to qubits: phase rotations and permutations give exact simulations within fixed-excitation subspaces while preserving the two-level input structure. We determine the exact, additive relative entropy of entanglement of the amplitude-damping resources and the exact regularized entanglement gain at fixed input population. We then derive explicit bounds for generalized amplitude damping, which includes thermal excitation. Combining the amplitude-damping gain with unital and entanglement-breaking channels gives a closed expression that recovers the zero-temperature result and vanishes throughout the entanglement-breaking region. It improves established benchmarks over broad parameter ranges, while the exact entanglement of the generalized resources remains open. Both results admit refinements under an average-excitation constraint. The bounds provide quantitative limits for adaptive communication through relaxing qubits and show how a sector simulation can remain useful beyond channels whose resource entanglement is known exactly.
\end{abstract}

\maketitle

\section{Introduction}
\label{sec:intro}

Amplitude damping describes a qubit that irreversibly loses an
excitation. It is also the action of optical pure loss on the vacuum
and one-photon states. The channel has long served as a test case
for quantum-capacity calculations~\cite{Wolf} and for error
correction adapted to excitation loss~\cite{Leung,Fletcher,Duan}.
Despite this simple physical interpretation,
determining how much quantum information or secret key can be
transmitted with two-way classical assistance remains difficult.
Classical feedback lets the parties distill shared entanglement
and then use it for teleportation~\cite{Teleport,BBPSSW,BDSW},
making the theory of mixed-state entanglement central to this
communication problem~\cite{HHHH}. The PLOB analysis uses the optical embedding to bound these rates and
also gives a tighter squashed-entanglement construction~\cite{PLOB}.
Other benchmarks follow from max-relative entropy~\cite{CMH,Rigovacca},
analytical evaluations of amplitude-damping bounds~\cite{ADseries},
and explicit distillation protocols~\cite{Abdelhadi}.

The optical embedding overlooks a useful restriction: each transmitted
mode contains at most one photon. Keeping that restriction throughout
the analysis leads here to a sharper analytic upper bound. The
improvement is strict at every nontrivial damping probability. At
$\gamma=1/2$, it lowers the optimized balanced squashed bound by
approximately $22.43\%$. Near complete damping, it is already within
about $0.46\%$ of the leading rate achieved by a known triple-rail code.

Channel simulation converts an adaptive communication protocol
into entanglement accounting on shared resources~\cite{PLOB,SimulationReview}.
Our approach groups several qubits according to their total number of
excitations. Within each such sector, the channel can be simulated
exactly by teleportation using a shared resource state. This extends
the sector method developed for bosonic channels~\cite{Bosonic,Noisy}
to a qubit setting. The distinction matters physically: general
optical mode mixing can create two photons in one mode, whereas
qubit phase rotations and permutations always preserve the allowed
local states. Those operations supply the symmetry needed for the
simulation~\cite{GBW}.

For amplitude damping, we also obtain an exact entanglement calculation. We identify
the optimal separable comparison states for the sector resources and
show that their relative entropy of entanglement is additive. This
also determines the regularized entanglement generated by a natural
family of entangled inputs. Consequently, the capacity bound cannot
be lowered simply by choosing better comparison states in the same
framework. Achieving that bound by a communication protocol remains
a separate question.

We next include excitation from a nonzero-temperature bath. Generalized
amplitude damping retains the phase and permutation symmetries needed
for sector simulation, but its resources contain overlapping loss and
excitation histories~\cite{KSW}. The amplitude-damping result supplies
a useful additional tool: a sharp upper bound on the increase of
regularized REE for an arbitrary input, including quantum memories.
Combining this gain with the known unital-channel REE and the
entanglement-breaking boundary produces an explicit GAD converse.
The resulting expression depends on both damping and bath population,
recovers the amplitude-damping result, and vanishes throughout the
entanglement-breaking region. It improves the evaluated established
benchmarks over broad parameter ranges; near part of the boundary,
the max-relative-entropy bound remains tighter. An average-excitation
constraint further refines both the ordinary and generalized results.

Section~\ref{sec:main} presents the amplitude-damping capacity bound
and its comparison with established benchmarks.
Section~\ref{sec:proof-sketch} sketches the proof through sector
simulation, exact resource entanglement, and adaptive entanglement
accounting, and states the associated optimality results.
Sections~\ref{sec:gad-channel}
and \ref{sec:gad-bounds} extend the construction and the capacity
bounds to generalized amplitude damping. The main text emphasizes the
physical results and comparisons, using established achievable
benchmarks. Appendix~\ref{sec:proof-structure} maps the proof dependencies
and the connection between the sketch and the detailed arguments.
Appendices~\ref{sec:simulation}--\ref{sec:comparisonproof}
give the amplitude-damping derivations; Appendices~\ref{sec:gad-proof}
and \ref{sec:gad-numerics} contain the GAD proofs and numerical
benchmark formulas.

\section{Capacity bound and comparison}
\label{sec:main}

\subsection{A relaxing qubit and its communication rates}

Let $\gamma$ be the decay probability and $\eta=1-\gamma$ the survival
probability. The amplitude-damping channel is
\begin{align}
 &\AD(\rho)=A_0\rho A_0^\dagger+A_1\rho A_1^\dagger,\nonumber\\
 A_0&=\proj0+\sqrt\eta\proj1,~~
 A_1=\sqrt\gamma\ket0\bra1.
 \label{eq:AD}
\end{align}
Thus $\ket1$ decays to $\ket0$ with probability $\gamma$, while the
ground state is unchanged. Identifying these basis states with the
one-photon state and vacuum makes $\AD$ the corresponding restriction
of bosonic pure loss.

Alice and Bob may use local quantum memories and unlimited authenticated
public two-way classical communication between channel uses. They
start without shared entanglement or secret key. We write $Q_2$ for
the quantum communication capacity, $D_2$ for entanglement distribution,
$P_2$ for private communication, and $K_2$ for secret-key agreement.
Rates are measured per physical channel use. Their standard relations
are $Q_2=D_2\leq P_2=K_2$. The private tasks are formulated through
private states~\cite{PrivateStates,PrivateParadigm}; the formal
protocol assumptions are stated
in Appendix~\ref{sec:adaptive}.

For the unrestricted qubit input, our main bound is
\begin{equation}
 \boxed{\Bsec(\gamma)=\log_2\!\left[1+(1-\gamma)
                    \gamma^{\gamma/(1-\gamma)}\right],}
 \label{eq:B}
\end{equation}
with continuous values $\Bsec(0)=1$ and $\Bsec(1)=0$.
It bounds all four two-way capacities.

The bound also incorporates an average-excitation constraint, in the
operational setting of energy-constrained two-way communication
developed in Ref.~\cite{DSW}. If $x_i$
is the excited-state population sent on use $i$, require, for $0\leq N\leq1$,
\begin{equation}
 x_i=\Tr(\proj1\rho^{(i)}_{A_i}),~~
 \frac1n\sum_{i=1}^n x_i\leq N.
 \label{eq:constraint}
\end{equation}
Here $\rho^{(i)}_{A_i}$ is the unconditional input marginal on use $i$;
the average includes all measurement outcomes and unsuccessful
branches. In the protocol notation of Sec.~\ref{sec:adaptive-main},
$\rho^{(i)}_{A_i}=(\rho_{i-1})_{A_i}$.
The function governing the bound is
\begin{equation}
 f_\gamma(x)=h_2((1-\gamma)x)-x h_2(\gamma),
 \label{eq:f}
\end{equation}
where $h_2$ is the binary entropy in bits. For $0<\gamma<1$, define the auxiliary quantity
$t_\gamma=\gamma^{\gamma/(1-\gamma)}$. The unique population that
maximizes $f_\gamma(x)$ over $0\leq x\leq 1$ is
\begin{equation}
 x_\gamma^\ast=\frac{t_\gamma}{1+(1-\gamma)t_\gamma}.\label{eq:xstar}
\end{equation}

The capacity statement, including the excitation constraint, is
\begin{equation}
 \begin{split}
 Q_2(\AD,N)&=D_2(\AD,N)
 \leq P_2(\AD,N)\\
 &=K_2(\AD,N)\leq\Bsec(\gamma,N),
 \end{split}
 \label{eq:main}
\end{equation}
where
\begin{equation}
 \Bsec(\gamma,N)=f_\gamma\bigl(\min\{N,x_\gamma^*\}\bigr).
 \label{eq:BN}
\end{equation}
For $N\geq x_\gamma^*$ this reduces to Eq.~\eqref{eq:B}; a smaller
allowed excitation lowers the bound. At zero damping the constrained
value is $h_2(\min\{N,1/2\})$, and at complete damping it is zero.
Appendix~\ref{sec:adaptive} gives the converse proof and finite-error
versions. No strong-converse assertion is needed here.

Preparing and measuring in the energy basis gives $f_\gamma$ a
classical interpretation. The induced binary channel is a Z-channel:
the ground state always gives output $0$, while an excitation gives
output $0$ with probability $\gamma$. If the input is excited with
probability $x$, the mutual information is
$I(X:Y)=h_2(\eta x)-x h_2(\gamma)=f_\gamma(x)$. Maximizing over
$0\leq x\leq 1$ gives the Shannon capacity of this Z-channel, equal
to $\Bsec(\gamma)$. Under the excitation constraint $x\leq N$, its
capacity is $\Bsec(\gamma,N)$. The quantum converse shows that these
same quantities upper-bound the corresponding two-way quantum and
private communication rates, even for coherent inputs and arbitrary
adaptive protocols.

\subsection{Established upper and lower bounds}
\label{sec:comparison}

For comparison, the optical embedding gives the PLOB REE
bound~\cite{PLOB}
\begin{equation}
 \UPLOB(\gamma)=\min\{1,-\log_2\gamma\},
 \label{eq:PLOB}
\end{equation}
and max-relative entropy gives~\cite{Rigovacca}
\begin{equation}
 \Umax(\gamma)=\log_2(2-\gamma).
 \label{eq:max}
\end{equation}
The balanced environmental-squashing construction in PLOB gives
\begin{align}
 \Usq(\gamma)&=\max_{0\leq x\leq1}s_\gamma(x),\nonumber\\
 s_\gamma(x)&=h_2\bigl((1-\gamma/2)x\bigr)-h_2(\gamma x/2).
 \label{eq:sq}
\end{align}
Here the input population is optimized, rather than fixed at $1/2$.
The label $\Usq$ denotes this particular squashing construction, not
the exact squashed entanglement of the channel. Our comparison is
analytic throughout the open damping interval:
\begin{equation}
 \Bsec(\gamma)<\min\{\UPLOB(\gamma),\Umax(\gamma),\Usq(\gamma)\}.
 \label{eq:strict}
\end{equation}
The proof is in Appendix~\ref{sec:comparisonproof}.

\begin{figure}[!t]
\vspace{-0.1cm}
\centering
\includegraphics[width=\columnwidth,trim={20bp 10bp 55bp 80bp},clip]
  {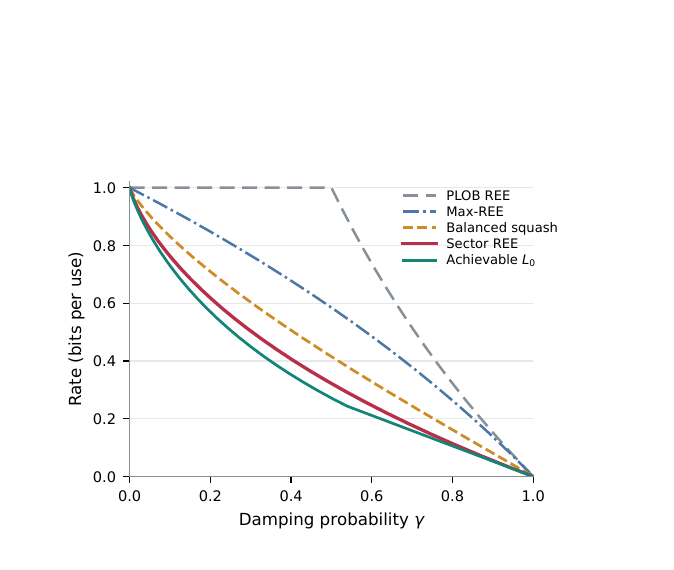}
\caption{Two-way capacity bounds for amplitude damping, over the full
damping range. The sector REE converse is compared with the PLOB REE,
max-relative-entropy, and optimized balanced squashed-entanglement
upper bounds in Eqs.~\eqref{eq:PLOB}--\eqref{eq:sq}. The established
achievable curve is $L_0=\max\{\RR,R_3\}$ from Eq.~\eqref{eq:lower}.
Our sector converse strictly improves all three displayed upper bounds
for any damping value in $0<\gamma<1$. The lower curve is a selected established benchmark, without an
optimality claim. Rates count every physical channel use.}
\label{fig:bounds}
\end{figure}

The established achievable reference combines reverse coherent
information~\cite{PGPBL,RCI,PLOB},
\begin{equation}
 \RR(\gamma)=\max_{0\leq x\leq1}
                         [h_2(x)-h_2(\gamma x)],
 \label{eq:RCI}
\end{equation}
with the triple-rail code~\cite{Abdelhadi}:
\begin{equation}
 \begin{split}
 R_3(\gamma)&=(1-\gamma)\frac{\log_2 3}{3},\\
 L_0(\gamma)&=\max\{\RR(\gamma),R_3(\gamma)\}.
 \end{split}
 \label{eq:lower}
\end{equation}
In triple rail, one excitation is shared coherently across three
channel uses. Survival preserves a three-dimensional entangled state;
loss is heralded by the vacuum. The rate therefore includes both
the survival probability and the three uses per attempt. Both terms
in $L_0$ are achievable for entanglement distribution, giving
\begin{equation}
 L_0(\gamma)\leq Q_2=D_2\leq P_2=K_2\leq\Bsec(\gamma).
 \label{eq:sandwich}
\end{equation}
Figure~\ref{fig:bounds} uses this established benchmark. More elaborate
distillation protocols are known~\cite{Abdelhadi}; the comparison here
keeps the transparent achievable reference $L_0$.

At $\gamma=1/2$, $x_\gamma^*=2/5$ and
$\Bsec=\log_2(5/4)$, compared with the balanced squashed value
$\log_2(4/3)$. Figure~\ref{fig:bounds} shows this $22.43\%$ reduction
relative to balanced squashing and the remaining gap to $L_0$.

\begin{figure}[!t]
\centering
\includegraphics[width=\columnwidth]{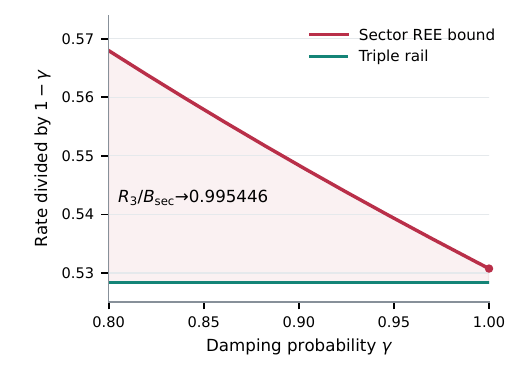}
\caption{High-damping comparison of the sector REE upper bound and the
established triple-rail rate, normalized by the survival probability
$1-\gamma$. The right endpoints are one-sided limits: all
unnormalized rates vanish at $\gamma=1$. Triple rail reaches more
than $99.54\%$ of the leading upper-bound coefficient, as quantified
by Eq.~\eqref{eq:ratio}.}
\label{fig:highdamping}
\end{figure}

Near complete damping, the sharper converse almost meets the
triple-rail rate. In terms of the small survival probability
$\eta=1-\gamma$,
\begin{equation}
 \Bsec(1-\eta)=\frac{\eta}{e\ln2}+O(\eta^2).
 \label{eq:asymptotic}
\end{equation}
The coefficient is $1/(e\ln2)=0.530737845\ldots$, while triple rail
has coefficient $0.528320834\ldots$. Thus
\begin{equation}
 \lim_{\gamma\uparrow1}\frac{R_3(\gamma)}{\Bsec(\gamma)}
     =\frac{e\ln3}{3}=0.995446\ldots.
 \label{eq:ratio}
\end{equation}
It therefore attains more than $99.54\%$ of the leading upper-bound
coefficient. Figure~\ref{fig:highdamping} resolves this small gap by dividing
the rates by the survival probability. The gap remains nonzero at this
order, so the comparison does not determine an exact capacity.

\section{Proof sketch and entanglement optimality}
\label{sec:proof-sketch}

The proof connects three physical ideas. Fixing the total excitation
number makes exact teleportation simulation possible; counting the
surviving excitations determines the resource entanglement; and the
adaptive converse framework converts that cost into a bound on
communication rates. We explain these steps and their optimality here.
Appendix~\ref{sec:proof-structure} gives a diagram of these dependencies.
The detailed qubit derivations are in
Appendices~\ref{sec:simulation}--\ref{sec:stateREE}; the established
bosonic converse tools are used through Refs.~\cite{Bosonic,Noisy}.
The resource upper estimate suffices for the capacity converse;
the exact entanglement calculation additionally establishes its
optimality within this accounting framework.

\subsection{Exact sector teleportation simulation}
\label{sec:physicalsimulation}

The useful symmetry appears when the total excitation number is fixed.
For $s$ qubits, the sector $\HH_{s,m}$ contains every configuration
with exactly $m$ excitations and has dimension $\binom{s}{m}$. For
example, the one-excitation sector of three qubits is spanned by
$\ket{100}$, $\ket{010}$, and $\ket{001}$. An arbitrary superposition
within this space still carries quantum information in the excitation
positions. Fixing their number does not measure those positions.

Write $\ket S$ for the configuration whose occupied sites form the
set $S$. The maximally entangled sector input is
\begin{equation}
 \ket{\Phi_{s,m}}=\frac{1}{\sqrt{\binom{s}{m}}}
                 \sum_{|S|=m}\ket S_R\ket S_{A'}.
 \label{eq:Phi}
\end{equation}
Alice can prepare both halves locally and transmit $A'$ through $s$
uses of the channel. This produces the shared resource
\begin{equation}
 J_{s,m}=(\id_R\otimes\AD^{\otimes s})(\Phi_{s,m}),
 \label{eq:sectorresource-main}
\end{equation}
where a pure-state symbol without a ket denotes its projector.
Alice retains $R$, while Bob holds the output block $B$.
This is the channel--state correspondence restricted to the
excitation sector~\cite{Jamiolkowski,Choi}.

Once $J_{s,m}$ is shared, the channel action on an input block $A$
supported in $\HH_{s,m}$ can be reproduced exactly by local
operations and classical communication (LOCC). We denote the
teleportation LOCC by $\mathcal T_{s,m}$. It acts on Alice's input
$A$ and resource share $R$, together with Bob's resource share $B$,
and returns the simulated channel output on Bob's side. Explicitly,
for every state $\rho_A$ supported in $\HH_{s,m}$,
\begin{equation}
 \AD^{\otimes s}(\rho_A)
   =\mathcal T_{s,m}\!\left(\rho_A\otimes J_{s,m}\right).
 \label{eq:sector-LOCC-main}
\end{equation}
Operationally, Alice makes a joint teleportation measurement on
$A R$, communicates its outcome, and Bob applies the corresponding
correction on $B$. The map $\mathcal T_{s,m}$ includes all measurement
outcomes and corrections, so the simulation is deterministic.
This construction uses the general relation between maximally
entangled measurements and teleportation
corrections~\cite{Teleport,Werner}.

Equation~\eqref{eq:sector-LOCC-main} is an identity of channels on the
sector, including their action on inputs entangled with external
quantum memories. In particular, for any memory $M$ and joint state
$\rho_{MA}$ whose $A$ support lies in $\HH_{s,m}$,
\begin{equation}
 \begin{split}
 &(\id_M\otimes\AD^{\otimes s})(\rho_{MA})\\
 &\hspace{1em}=
 (\id_M\otimes\mathcal T_{s,m})
       \!\left(\rho_{MA}\otimes J_{s,m}\right).
 \end{split}
 \label{eq:sector-LOCC-memory}
\end{equation}
The memory may include systems held by either party.

The correction unitaries require only permutations of the qubits
and local phase rotations. These operations preserve each excitation
sector, and amplitude damping respects their symmetry. Together they
act irreducibly within $\HH_{s,m}$, supplying the symmetry needed
for the covariant teleportation construction~\cite{GBW}.
Appendix~\ref{sec:simulation} verifies this condition and gives the
measurement defining $\mathcal T_{s,m}$ explicitly.

The construction provides a reusable simulation tool for tensor powers
of phase-covariant qubit channels, including generalized amplitude
damping, dephasing, and depolarization~\cite{Ruskai,KSW}. In each case,
it gives an exact simulation on every fixed-excitation input sector,
using the corresponding channel output of $\Phi_{s,m}$ as the
resource. Obtaining an improved capacity bound then requires a
sufficiently sharp, channel-dependent estimate of the resource
entanglement. We evaluate this entanglement exactly for amplitude
damping in the next subsection. The extension from fixed-sector
inputs to arbitrary adaptive protocols is explained in
Sec.~\ref{sec:adaptive-main}.

\subsection{Exact resource entanglement}
\label{sec:exact-main}

The resource cost is measured by the relative entropy of entanglement
(REE), $\ER$: the minimum quantum relative entropy from the shared
state to a separable state~\cite{Vedral,VP,PlenioVirmani}. It cannot increase under
local operations and classical communication, so it limits what the
simulated channel can distribute. Its regularization, $\ERinf$, is
the REE per copy when arbitrarily many identical copies are considered
together. Detailed mathematical definitions are provided in
Appendix~\ref{sec:resource}.

For amplitude damping, Bob can count the surviving excitations without
resolving their positions. If the input contains $m$ excitations, the
survivor count $K$ is binomial. The resource separates into conditional
states $\tau_{s,m,k}$:
\begin{equation}
 J_{s,m}=\sum_{k=0}^{m}p_{m,k}\tau_{s,m,k},~~
 p_{m,k}=\binom{m}{k}\eta^k\gamma^{m-k}.
 \label{eq:resourceblocks}
\end{equation}
The states with different $k$ can be distinguished locally by Bob.
Our exact conditional result is
\begin{equation}
 \boxed{\ER(\tau_{s,m,k})=\ERinf(\tau_{s,m,k})
               =\log_2\frac{\binom{s}{k}}{\binom{m}{k}}.}
 \label{eq:REEexact}
\end{equation}
The two binomial coefficients count possible survivor configurations:
$\binom{s}{k}$ over the whole block, and $\binom{m}{k}$ within a
fixed occupied input set. At $k=m$, no excitation is lost and the
answer is the maximal sector entanglement $\log_2\binom{s}{m}$.
At $k=0$, no entanglement survives.

Because the survivor count is locally accessible, the unconditional
REE is its average:
\begin{equation}
 F_s(m):=\ER(J_{s,m})
   =\EE_{K\sim\Bin(m,\eta)}\log_2\frac{\binom{s}{K}}{\binom{m}{K}}.
 \label{eq:F}
\end{equation}
These REEs are additive, even for products of different sector
resources. In particular, $\ERinf(J_{s,m})=F_s(m)$. For every finite
block, $F_s(m)\leq s f_\gamma(m/s)$. Moreover, as $s\to\infty$ at
fixed excitation density,
\begin{equation}
 \frac{F_s(m_s)}{s}\longrightarrow f_\gamma(x),~~
 \frac{m_s}{s}\longrightarrow x.
 \label{eq:resource-density-main}
\end{equation}
Thus $f_\gamma(x)$ is the exact asymptotic resource cost per channel
use at input population $x$. This gives the physical origin of the
function governing the capacity bound in Eq.~\eqref{eq:BN}.

Equation~\eqref{eq:REEexact} follows from an explicit separable
comparison state, constructed by averaging products of local
fixed-excitation superpositions over phases. We prove that it is
optimal among \emph{all} separable states, and that joint comparisons
across resources give no further reduction. Appendix~\ref{sec:resource}
contains the construction and proofs. The resource entanglement
density entering the converse is therefore fixed by this exact
optimization over separable states.

\subsection{Adaptive converse and optimality}
\label{sec:adaptive-main}

Consider an $n$-use adaptive protocol
$\mathcal P_n=(\rho_0,\Lambda_1,\ldots,\Lambda_n)$.
The initial state $\rho_0$ includes the first transmitted qubit $A_1$
and is prepared by LOCC without shared entanglement or secret key. Each LOCC map $\Lambda_i$ includes the public
measurement outcomes and their feedback; for $i<n$, it also prepares
the next transmitted qubit. Let $\rho_i$ denote the joint state of
Alice's and Bob's retained registers after the $i$th transmission
and the following LOCC. For $i=1,\ldots,n$,
\begin{equation}
 \begin{split}
 \rho_i&=\Lambda_i\!\left[
    (\id_{\mathrm{mem}}\otimes\AD)(\rho_{i-1})\right],\\
 \zeta_n&:=\rho_n,
 \end{split}
 \label{eq:protocol-state-main}
\end{equation}
where $\AD$ acts on Alice's transmitted qubit $A_i$ and
$\id_{\mathrm{mem}}$ leaves all other registers unchanged.
Thus $\zeta_n$ is the final retained state of the protocol.
For key generation, we use the input-preserving retained-state
construction of Appendix~\ref{sec:adaptive}, so $\zeta_n$ includes
the local shield systems before the final key measurement.

All states here are unconditional: they include every outcome and
unsuccessful branch. The population transmitted on use $i$ is
$x_i=\Tr[\proj1(\rho_{i-1})_{A_i}]$, with
$n^{-1}\sum_i x_i\leq N$ as in Eq.~\eqref{eq:constraint}.

The converse follows by applying the sector construction to $r$
independent replicas of this entire protocol. At each round, group
the $r$ corresponding transmitted qubits. Combining their
fixed-sector simulations $\mathcal T_{r,m}$, the resource costs
$F_r(m)$, and the established sector-coherence
estimate~\cite{Bosonic,Noisy} gives
\begin{equation}
 \ER(\zeta_n^{\otimes r})
 \leq r\sum_{i=1}^{n}f_\gamma(x_i)+n\log_2(r+1).
 \label{eq:replica-main}
\end{equation}
This is Eq.~\eqref{eq:replica} below, obtained by applying the block
estimate \eqref{eq:block} to each replicated round.
The logarithmic term accounts for coherence between excitation
sectors; the actual protocol inputs need not be measured or dephased.
Correlations between successive transmissions within each adaptive
protocol, including those carried by quantum memories, remain
unrestricted.

Dividing by $r$ and taking $r\to\infty$ at fixed $n$ removes this
overhead and yields
\begin{equation}
 \ERinf(\zeta_n)
 \leq\sum_{i=1}^{n}f_\gamma(x_i)
 \leq n\Bsec(\gamma,N).
 \label{eq:adaptive-main}
\end{equation}
Concavity bounds the sum by $n f_\gamma(\bar x)$, where
$\bar x=n^{-1}\sum_i x_i\leq N$. Maximizing over the allowed
mean population gives
$\Bsec(\gamma,N)=f_\gamma(\min\{N,x_\gamma^*\})$.

Finally, entanglement and privacy-test estimates connect
$\ERinf(\zeta_n)$ to a target of dimension $d_n$. As the
communication error vanishes, they bound the achievable rate
$\log_2 d_n/n$ by $\Bsec(\gamma,N)$, establishing
Eq.~\eqref{eq:main}. Without an excitation constraint, the
optimization gives Eq.~\eqref{eq:B}.
Appendix~\ref{sec:adaptive} provides the precise finite-error
inequalities and the treatment of arbitrary memories. The resulting
bounds are weak converses; no strong-converse conclusion is used.

The same construction also determines how much regularized
entanglement a single transmission can generate. Prepare a pair with
excited-state population $x$ and send one half:
\begin{align}
 \ket{\varphi_x}_{RA}&=\sqrt{1-x}\ket{00}+\sqrt{x}\ket{11},\nonumber\\
 \Omega_{\gamma,x}&=(\id_R\otimes\AD)(\varphi_x).
 \label{eq:Omegax}
\end{align}
The exact regularized output entanglement is
\begin{equation}
 \boxed{\ERinf(\Omega_{\gamma,x})=f_\gamma(x).}
 \label{eq:OmegaxREE}
\end{equation}
Moreover, $f_\gamma(x)$ is the largest possible increase of regularized
REE for one use at that population. This is an exact evaluation
of a regularized amortized entanglement gain. It allows
correlated quantum memories and finite initial REE. Preparing
$\varphi_x$ entirely in Alice's laboratory before transmission
attains the increase. Appendix~\ref{sec:stateREE} proves both
statements. This entanglement-gain theorem also supplies the input
for the generalized amplitude-damping converse in Section~\ref{sec:gad-bounds}.

This identifies the limit of the present converse method. Optimizing
the separable comparisons, including collective comparisons over many
resources, cannot lower Eq.~\eqref{eq:B}. A tighter capacity bound
would need additional information beyond this regularized-REE
accounting.

The fixed-excitation resource states have additive REE. For the probe
output $\Omega_{\gamma,x}$ with $0<\gamma,x<1$, however, the single-copy
REE strictly exceeds the regularized value in Eq.~\eqref{eq:OmegaxREE}
except at $x=1/(2-\gamma)$, where the no-decay component is maximally
entangled and the two values coincide
(see Corollary~\ref{cor:probe-strict} in Appendix~\ref{sec:stateREE}). Since $t_\gamma<1$,
Eq.~\eqref{eq:xstar} gives $x_\gamma^\ast<1/(2-\gamma)$. Thus the
inequality is strict even at the population that maximizes the
converse bound. Neither Eq.~\eqref{eq:OmegaxREE} nor the optimality
of the sector comparison states identifies REE with distillable
entanglement.

\section{Thermal excitation and sector simulation}
\label{sec:gad-channel}

A bath at nonzero temperature can excite a qubit as well as relax it.
The corresponding generalized amplitude-damping (GAD) channel is
\begin{align}
 \GAD&=(1-q)\AD+q\AD^\uparrow,\nonumber\\
 \AD^\uparrow(\rho)&=X\AD(X\rho X)X,
 \label{eq:gad-channel}
\end{align}
where $X$ is the Pauli bit flip and $q$ is the equilibrium excitation
probability. An input population $x$ is mapped to
\begin{equation}
 y=(1-\gamma)x+\gamma q,
 \label{eq:gad-population}
\end{equation}
while its coherence is multiplied by $\sqrt{1-\gamma}$. The convention
agrees with the bath parameter $N$ of Ref.~\cite{KSW}; here $N$ remains
reserved for the input constraint. The interval $0\leq q\leq1/2$
describes a bath without population inversion. We give formulas for
the full interval $0\leq q\leq1$, using the unitary equivalence under
$q\leftrightarrow1-q$. Ordinary amplitude damping is recovered at
$q=0$, and $q=1$ gives its excitation counterpart.

The sector teleportation construction survives this extension.
The GAD channel respects independent phase rotations, and identical
channel factors respect qubit permutations. The same corrections
therefore simulate $\GAD^{\otimes s}$ exactly on $\HH_{s,m}$, using
\begin{equation}
 J_{s,m}^{\gamma,q}
   =(\id_R\otimes\GAD^{\otimes s})(\Phi_{s,m}).
 \label{eq:gad-resource}
\end{equation}
With the same teleportation LOCC as in
Eq.~\eqref{eq:sector-LOCC-main}, the simulation reads
\begin{equation}
 \GAD^{\otimes s}(\rho_A)
 =\mathcal T_{s,m}\!\left(\rho_A\otimes J_{s,m}^{\gamma,q}\right),
 \label{eq:gad-sector-LOCC}
\end{equation}
for $\rho_A$ supported in $\HH_{s,m}$. The identity includes arbitrary
external memories, as in Eq.~\eqref{eq:sector-LOCC-memory}.
Appendix~\ref{sec:gad-proof} applies the general simulation statement
of Appendix~\ref{sec:simulation} to this channel.

Thermal excitation changes the entanglement calculation. An output
excitation count can now arise from several combinations of losses
and excitations, so it does not identify the same conditional
resources as in the zero-temperature case. We obtain a converse by
combining the amplitude-damping entanglement-gain theorem with two
tractable members of the GAD family: the unital channel and an
entanglement-breaking channel. This gives an explicit extension
without requiring an exact REE formula for
$J_{s,m}^{\gamma,q}$.

\begin{figure*}[!t]
\centering
\includegraphics[width=0.9\textwidth]{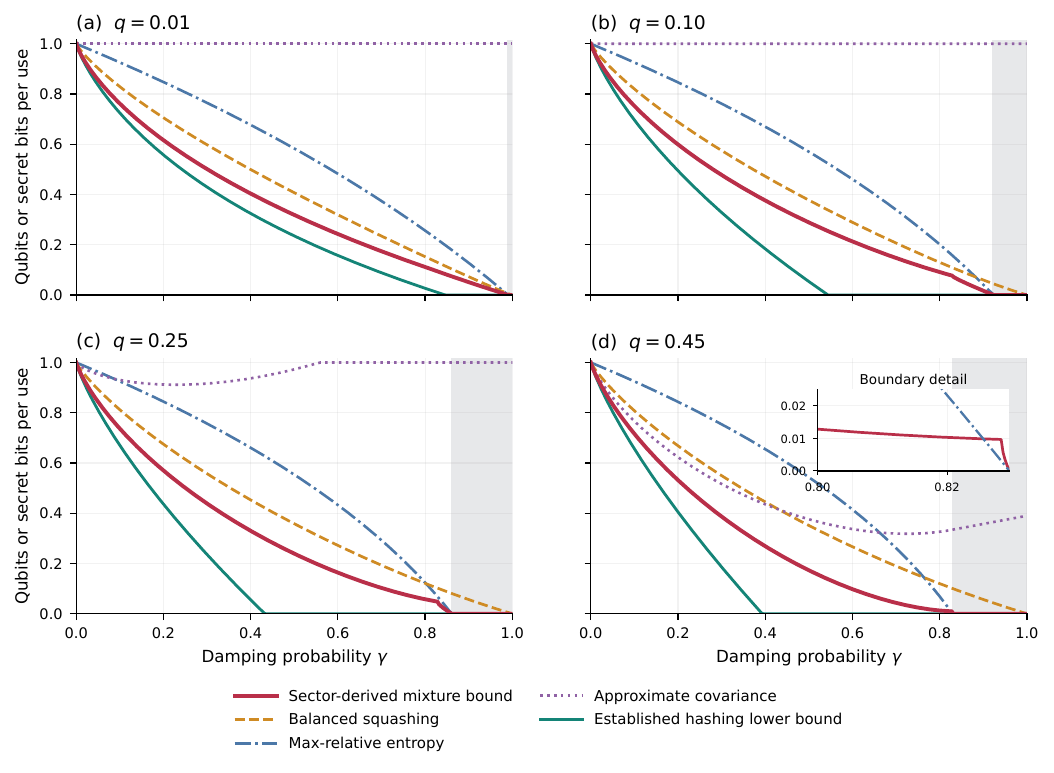}
\caption{Two-way capacity bounds for generalized amplitude damping at
four equilibrium excitation probabilities. The red curve is the
sector-derived mixture converse in Eq.~\eqref{eq:gad-mix}. The
other upper curves are the balanced-squashing, max-relative-entropy,
and approximate-covariance bounds of Ref.~\cite{KSW}; the green
curve is the established diagonal-input hashing benchmark.
The two balanced squashing constructions are separately optimized
over the input population and their smaller value is plotted.
Gray shading denotes the entanglement-breaking region, where all
four capacities vanish. The inset in panel (d) resolves a region
where max-relative entropy improves the red bound. No numerical
ordering is claimed outside the displayed comparisons.}
\label{fig:gad-bounds}
\end{figure*}

\section{Two-way capacity bounds for generalized damping}
\label{sec:gad-bounds}

\subsection{An explicit bound depending on damping and bath excitation}

The unital member, $\mathcal G_{\gamma,1/2}$, is a Pauli channel. Its
Choi state is Bell diagonal and has the known REE~\cite{PLOB,KSW}
\begin{align}
 \lambda_\gamma&=\frac{(1+\sqrt{1-\gamma})^2}{4},\nonumber\\
 P_\gamma&=
 \begin{cases}
 1-h_2(\lambda_\gamma),&\lambda_\gamma>1/2,\\
 0,&\lambda_\gamma\leq1/2.
 \end{cases}
 \label{eq:gad-pauli}
\end{align}
Its exact teleportation simulation bounds its entanglement gain by
$P_\gamma$. A second useful reference is the boundary of the known
entanglement-breaking region~\cite{KSW},
\begin{equation}
 \gamma^2q(1-q)\geq1-\gamma.
 \label{eq:gad-eb}
\end{equation}
Such a channel cannot create entanglement or secret key between
parties who initially share neither.

To state the result, set $r=\min\{q,1-q\}$ and
\begin{align}
 \gamma_c&=2(\sqrt2-1),\nonumber\\
 r_c(\gamma)&=\frac{1-\sqrt{1-4(1-\gamma)/\gamma^2}}{2},
 ~~\gamma_c\leq\gamma<1.
 \label{eq:gad-threshold}
\end{align}
With the same operational assumptions as in the amplitude-damping
case, our GAD converse is
\begin{equation}
 \begin{split}
 Q_2(\GAD)&=D_2(\GAD)\leq P_2(\GAD)\\
 &=K_2(\GAD)\leq\Umix(\gamma,q),
 \end{split}
 \label{eq:gad-capacities}
\end{equation}
where
\begin{equation}
 \boxed{
 \Umix(\gamma,q)=
 \begin{cases}
 (1-2r)\Bsec(\gamma)+2rP_\gamma,
                         &\gamma<\gamma_c,\\[2pt]
 [1-r/r_c(\gamma)]_+\Bsec(\gamma),
                         &\gamma_c\leq\gamma<1,\\[2pt]
 0,                      &\gamma=1.
 \end{cases}}
 \label{eq:gad-mix}
\end{equation}
Here $[z]_+=\max\{z,0\}$ and $\Bsec$ is the amplitude-damping bound
in Eq.~\eqref{eq:B}. Equation~\eqref{eq:gad-mix} recovers
$\Bsec$ at $q=0,1$, recovers $P_\gamma$ at $q=1/2$, and vanishes
throughout the entanglement-breaking region. These consistency
limits follow directly from the expression.

The physical decomposition behind the first branch is particularly
simple. For $q\leq1/2$,
\begin{equation}
 \GAD=(1-2q)\AD+2q\mathcal G_{\gamma,1/2}.
 \label{eq:gad-unitalmix-main}
\end{equation}
The increase of regularized REE is bounded by the corresponding
average of component gains. For $\gamma_c\leq\gamma<1$ and
$q<r_c(\gamma)$, decomposing into $\AD$ and
$\mathcal G_{\gamma,r_c(\gamma)}$ improves the estimate because the
latter channel is entanglement breaking. For
$r_c(\gamma)\leq q\leq1/2$, the original channel is already
entanglement breaking. The range $q>1/2$ follows by the bit-flip
equivalence in Eq.~\eqref{eq:gad-reflection}. The proof uses
the amortized gain established in Appendix~\ref{sec:stateREE} and
convexity of regularized REE; it does not assume convexity of channel
capacities. Appendix~\ref{sec:gad-proof} gives the complete argument.

\subsection{Average-excitation refinement}
\label{sec:gad-energy-main}

The same reasoning retains the population entering every channel
use. Define
\begin{equation}
 F_{\gamma,q}(x)=(1-q)f_\gamma(x)+qf_\gamma(1-x),
 \label{eq:gad-branch-profile}
\end{equation}
where $f_\gamma$ is given in Eq.~\eqref{eq:f}. Let
$z_q(x)=x$ for $q\leq1/2$ and $z_q(x)=1-x$ for $q>1/2$, and put
\begin{equation}
 H_{\gamma,q}(x)=
 \begin{cases}
 (1-2r)f_\gamma(z_q(x))+2rP_\gamma,
                           &\gamma<\gamma_c,\\[2pt]
 [1-r/r_c(\gamma)]_+f_\gamma(z_q(x)),
                           &\gamma_c\leq\gamma<1,\\[2pt]
 0,                        &\gamma=1.
 \end{cases}
 \label{eq:gad-hprofile}
\end{equation}
Then each of the four capacities, denoted here by $C_2$, obeys
\begin{equation}
 \boxed{
 C_2(\GAD,N)\leq
 \max_{0\leq x\leq N}
       \min\{F_{\gamma,q}(x),H_{\gamma,q}(x)\},}
 \label{eq:gad-energy}
\end{equation}
for $0\leq N\leq1$ under Eq.~\eqref{eq:constraint}.
Only a one-dimensional optimization is required. The result
reduces to Eq.~\eqref{eq:BN} at $q=0$ and is zero at $N=0$ for
every bath population. Bath-induced excitation at the receiver
therefore does not replace the sender's input resource. The
unconstrained symmetry $q\leftrightarrow1-q$ does not imply the
same symmetry at a fixed input budget: the corresponding bit flip
also changes $x$ to $1-x$.

\subsection{Comparison with established bounds}
\label{sec:gad-comparison}

Figure~\ref{fig:gad-bounds} compares the closed expression
\eqref{eq:gad-mix} with the max-relative-entropy bound, the smaller
of the two balanced environmental-squashing bounds, and the
approximate-covariance bound in Ref.~\cite{KSW}. The lower curve
uses the established hashing rates in both communication
directions~\cite{DW,HOW,PGPBL,RCI}: for a Schmidt-diagonal input,
it optimizes the larger of the coherent and reverse coherent
information, together with zero. Appendix~\ref{sec:gad-numerics}
specifies these benchmarks and their numerical evaluation.

The comparison shows improvement over broad ranges of damping and
bath population. The ordering is not uniform: near part of the
entanglement-breaking boundary, the max-relative-entropy bound is
tighter, as resolved by the inset. The useful combined converse is
the minimum of Eq.~\eqref{eq:gad-mix} and the independent established
upper bounds. The figure reports the selected benchmark
constructions; it does not establish optimality over arbitrary
squashing channels or all converse methods.

The distinction from the amplitude-damping case is substantive. For amplitude
damping, the resource REE and its additivity are exact, so better
separable comparisons cannot improve that converse. For GAD, the
mixture estimate is an upper bound on the entanglement gain, and the
exact REE of the unflagged resources in Eq.~\eqref{eq:gad-resource}
remains open. Their direct optimization could therefore improve the
generalized result.

\section{Conclusions}
\label{sec:conclusion}

For amplitude damping, retaining the qubit restriction of pure loss
through fixed-excitation sectors yields a closed upper bound on all
four two-way-assisted capacities. It strictly improves the PLOB REE,
analytic max-relative-entropy, and optimized balanced squashed bounds
throughout the nontrivial damping interval. Near complete damping,
the established triple-rail rate attains more than $99.54\%$ of its
leading coefficient. Exact resource REE, additivity, and the sharp
regularized entanglement gain establish the limit of separable-state
optimization within this construction. Whether an achievable
protocol closes the remaining gap is a separate problem.

For generalized amplitude damping, phase and permutation covariance
still give exact sector simulation. Combining the amplitude-damping
gain with the unital member and the entanglement-breaking boundary
of the generalized family yields a two-parameter converse. It
recovers the ordinary damping result and the known unital bound,
vanishes throughout the entanglement-breaking region, and improves
the evaluated benchmarks over broad ranges. Its combination with
independent bounds remains useful where their ordering changes.
The exact entanglement and additivity of generalized sector
resources are not determined by this extension; their direct
optimization offers a route to further improvement.

Both cases retain the unconditional mean input-excitation constraint
throughout adaptive communication. The established bosonic
pinching, memory, and retained-state tools provide the operational
framework, while the qubit symmetry and channel-specific
entanglement estimates supply the new inputs. The progression from
an exact resource calculation for amplitude damping to a mixture
estimate for generalized damping demonstrates two ways to use the
same simulation structure for physically constrained converse
bounds. Further work can address generalized resource entanglement,
stronger achievable rates, and finite-blocklength refinements.

\par\medskip
\noindent\textbf{AI statement.}
OpenAI's ChatGPT (GPT-6 and GPT-6.2 Astra models) together with Anthropic's Claude (Opus 5.5) were used during the preparation of this manuscript to generate candidate mathematical
derivations and proofs, assist with their development, and improve the presentation of the manuscript. The tools were directed through
author-specified prompts concerning the mathematical assumptions,
derivations, and statements under consideration. All AI-assisted
mathematical results, derivations, and proofs were independently
examined, verified, and revised as necessary by the author, who assumes responsibility for the accuracy and scientific content of the work.

\appendix

\clearpage
\onecolumngrid
\section{Structure of the proof}
\label{sec:proof-structure}

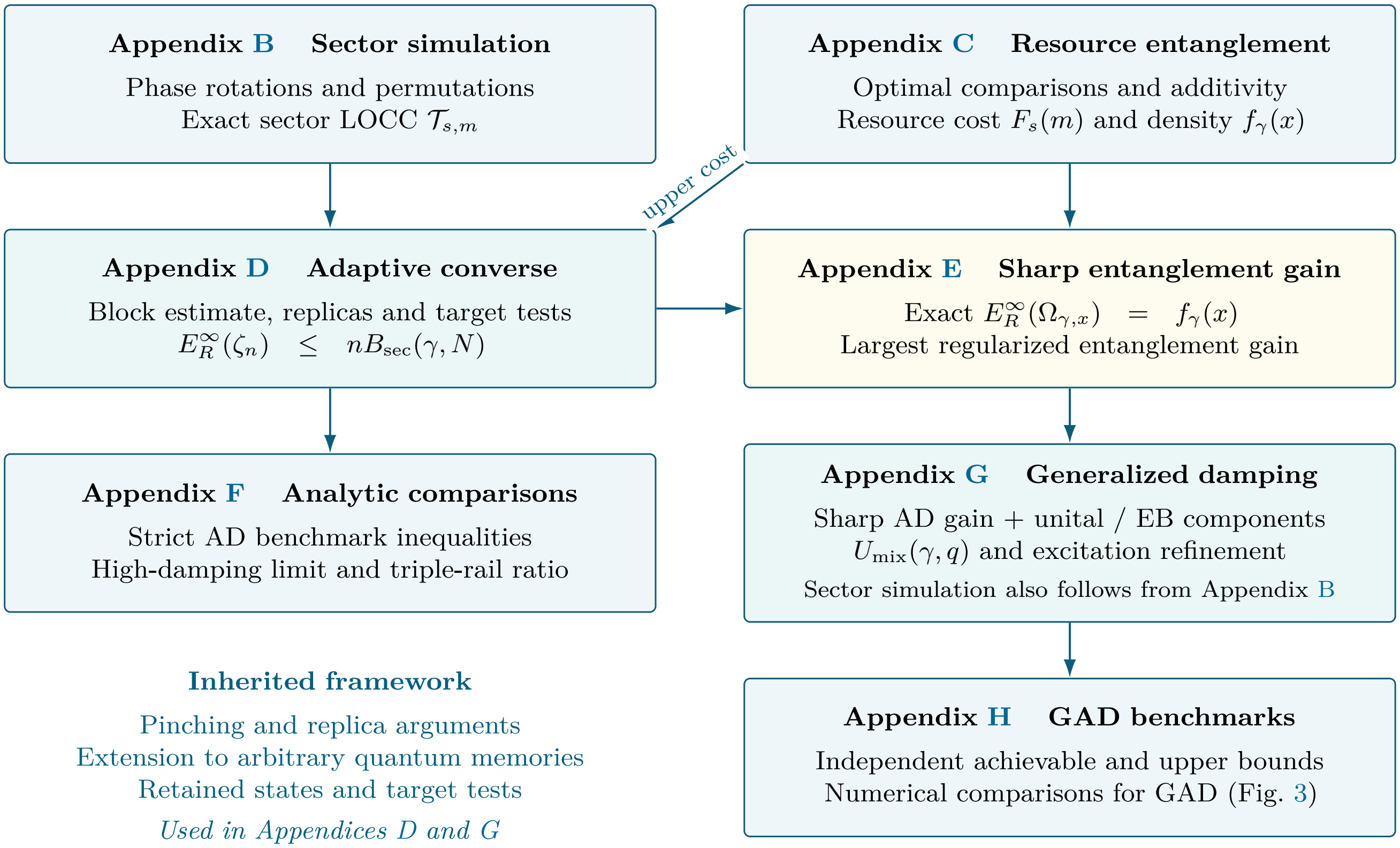
\begin{figure}[!ht]
\centering
\resizebox{0.96\textwidth}{!}{%
\begin{tikzpicture}[
  proofbox/.style={
    rectangle, rounded corners=2pt,
    draw=MidnightBlue!65!black, line width=0.55pt,
    fill=MidnightBlue!4,
    text width=7.2cm, minimum height=1.9cm,
    inner xsep=9pt, inner ysep=7pt,
    align=center, font=\fontsize{9}{11}\selectfont
  },
  converse/.style={proofbox, fill=TealBlue!8},
  exactgain/.style={proofbox, fill=Goldenrod!9},
  proofarrow/.style={
    -{Latex[length=2.2mm,width=1.5mm]},
    draw=MidnightBlue!70!black, line width=0.7pt
  },
  edgelabel/.style={
    font=\fontsize{8}{10}\selectfont,
    text=MidnightBlue!75!black, fill=white,
    inner sep=2pt
  }
]
\node[proofbox] (simulation) at (0,0) {
  \textbf{Appendix~\ref{sec:simulation} ~~ Sector simulation}\\[4pt]
  Phase rotations and permutations\\
  Exact sector LOCC $\mathcal T_{s,m}$
};
\node[proofbox] (resource) at (8.9,0) {
  \textbf{Appendix~\ref{sec:resource} ~~ Resource entanglement}\\[4pt]
  Optimal comparisons and additivity\\
  Resource cost $F_s(m)$ and density $f_\gamma(x)$
};
\node[converse] (adaptive) at (0,-2.7) {
  \textbf{Appendix~\ref{sec:adaptive} ~~ Adaptive converse}\\[4pt]
  Block estimate, replicas and target tests\\
  $\ERi(\zeta_n)\leq n\Bsec(\gamma,N)$
};
\node[exactgain] (gain) at (8.9,-2.7) {
  \textbf{Appendix~\ref{sec:stateREE} ~~ Sharp entanglement gain}\\[4pt]
  Exact $\ERi(\Omega_{\gamma,x})=f_\gamma(x)$\\
  Largest regularized entanglement gain
};
\node[proofbox] (comparisons) at (0,-5.4) {
  \textbf{Appendix~\ref{sec:comparisonproof} ~~ Analytic comparisons}\\[4pt]
  Strict AD benchmark inequalities\\
  High-damping limit and triple-rail ratio
};
\node[converse] (gad) at (8.9,-5.4) {
  \textbf{Appendix~\ref{sec:gad-proof} ~~ Generalized damping}\\[4pt]
  Sharp AD gain + unital / EB components\\
  $\Umix(\gamma,q)$ and excitation refinement\\[2pt]
  {\fontsize{8}{10}\selectfont
   Sector simulation also follows from Appendix~\ref{sec:simulation}}
};
\node[proofbox] (benchmarks) at (8.9,-8.1) {
  \textbf{Appendix~\ref{sec:gad-numerics} ~~ GAD benchmarks}\\[4pt]
  Independent achievable and upper bounds\\
  Numerical comparisons for GAD (Fig.~\ref{fig:gad-bounds})
};
\node[
  text width=7.2cm, align=center,
  font=\fontsize{9}{11}\selectfont,
  text=MidnightBlue!75!black
] (framework) at (0,-8.1) {
  \textbf{Inherited framework}\\[4pt]
  Pinching and replica arguments\\
  Extension to arbitrary quantum memories\\
  Retained states and target tests\\[3pt]
  \textit{Used in Appendices~\ref{sec:adaptive} and~\ref{sec:gad-proof}}
};
\draw[proofarrow] (simulation.south) -- (adaptive.north);
\draw[proofarrow] (resource.south west) --
  node[edgelabel, above, sloped] {upper cost} (adaptive.north east);
\draw[proofarrow] (resource.south) -- (gain.north);
\draw[proofarrow] (adaptive.east) -- (gain.west);
\draw[proofarrow] (adaptive.south) -- (comparisons.north);
\draw[proofarrow] (gain.south) -- (gad.north);
\draw[proofarrow] (gad.south) -- (benchmarks.north);
\end{tikzpicture}%
}
\caption{Main dependencies of the proof. The resource upper estimate in Appendix~\ref{sec:resource}
and sector simulation in Appendix~\ref{sec:simulation} suffice for the
adaptive converse in Appendix~\ref{sec:adaptive}. The exact part of the
resource calculation, together with the block estimate, establishes
the sharp entanglement gain in Appendix~\ref{sec:stateREE}. That gain
supports the GAD mixture converse in Appendix~\ref{sec:gad-proof}.
The GAD sector simulation also uses Appendix~\ref{sec:simulation};
an exact GAD resource-entanglement formula is not required.}
\label{fig:proof-structure}
\end{figure}

Figure~\ref{fig:proof-structure} organizes the detailed arguments behind
the proof sketch in Sec.~\ref{sec:proof-sketch}. The amplitude-damping
converse combines an exact simulation on each excitation sector with
an upper estimate of its resource cost. Exact separable optimality and
additivity then identify the limit of this entanglement accounting and
supply the sharp gain used for generalized amplitude damping (GAD).
The general pinching, memory-extension, replica, and retained-state
tools are imported from Refs.~\cite{Bosonic,Noisy}; the appendices below
give the qubit-specific inputs and their application.

\noindent\textbf{Amplitude-damping converse and optimality.}
Appendix~\ref{sec:simulation} constructs the teleportation LOCC
$\mathcal T_{s,m}$ from the irreducible phase--permutation action
(Lemma~\ref{lem:irrep} and Proposition~\ref{prop:simulation}).
Appendix~\ref{sec:resource} gives the comparison state, proves its
optimality and additivity, and establishes
$F_s(m)\leq s f_\gamma(m/s)$ with exact limiting density
(Theorem~\ref{thm:optimal} and Lemma~\ref{lem:density}).
Appendix~\ref{sec:adaptive} combines the simulation and this upper
estimate with the inherited coherence control to obtain
Eq.~\eqref{eq:block}. Replicas of the entire adaptive protocol remove
the logarithmic overhead, and the retained-state tests give
Theorem~\ref{thm:main} and Corollary~\ref{cor:finite}. Appendix~\ref{sec:stateREE} uses the exact resource density and the
block estimate to prove Theorem~\ref{thm:regularized}. This theorem
establishes the sharp regularized-REE gain at fixed input population.
The amplitude-damping capacity bound requires only the upper estimate
on this gain. Corollary~\ref{cor:probe-strict} in the same appendix
characterizes when the probe's single-copy REE strictly exceeds its
regularized value. The GAD converse uses the gain bound from the
theorem and does not depend on the corollary.
Appendix~\ref{sec:comparisonproof} establishes the strict benchmark
comparisons and the high-damping limit.

\medskip
\noindent\textbf{GAD extension and comparisons.}
Appendix~\ref{sec:gad-proof} first applies the same sector LOCC to GAD.
Its capacity argument combines the amplitude-damping gain from
Appendix~\ref{sec:stateREE}, the unital-channel estimate, and the
entanglement-breaking boundary through Lemma~\ref{lem:gad-convexity}.
Telescoping these gains over the retained adaptive protocol and
applying the tests of Appendix~\ref{sec:adaptive} gives
Eqs.~\eqref{eq:gad-capacities} and \eqref{eq:gad-energy}.
Appendix~\ref{sec:gad-numerics} specifies the independent benchmark
formulas used in Fig.~\ref{fig:gad-bounds}. Neither the exact GAD
resource REE nor capacity achievability is asserted by this route.

\clearpage
\twocolumngrid

\section{Sector symmetry and exact teleportation simulation}
\label{sec:simulation}

We prove the sector telesimulation used in Sec.~\ref{sec:physicalsimulation}.
Only the qubit symmetry and its substitution into the established
covariant teleportation construction are required. For $s$ qubits define the total excitation operator and its sectors by
\begin{equation}
 \begin{split}
 \widehat M_s&=\sum_{j=1}^s\proj1_j,\\
 \HH_{s,m}&=\operatorname{span}\{\ket S:S\subseteq[s],\ |S|=m\},
 \end{split}
 \label{eq:sector}
\end{equation}
where $[s]=\{1,\ldots,s\}$ and $\ket S$ has excitations precisely at
the sites in $S$. The projector onto this space is $\Pi_{s,m}$ and its
dimension is $d_{s,m}=\binom{s}{m}$, for $0\leq m\leq s$. For example, when $s=4$ and $m=2$,
\begin{align}
\HH_{4,2}
=
\operatorname{span}\{
&\ket{1100},
\ket{1010},
\ket{1001}, \nonumber \\
&\ket{0110},
\ket{0101},
\ket{0011}
\},
\end{align}
corresponding to the six subsets of $[4]$ with cardinality two. For
instance, if $S=\{1,3\}$, then
\begin{equation}
\ket S=\ket{1010}.
\end{equation}

The maximally entangled input is $\Phi_{s,m}$ from
Eq.~\eqref{eq:Phi}. We now specify its symmetry group.

Let $\omega_3=e^{2\pi i/3}$ and consider
\begin{equation}
 G_s=\ZZ_3^s\rtimes S_s,~~
 D_{\bm t}=\bigotimes_{j=1}^s
       \bigl(\proj0+\omega_3^{t_j}\proj1\bigr),
 \label{eq:group}
\end{equation}
where $\rtimes$ is the semidirect product associated with the
natural action of $S_s$ on $\ZZ_3^s$ by permutation of coordinates.

For $g=(\bm t,\pi)$ let $U_g=D_{\bm t} P_\pi$, where $P_\pi$ permutes
the qubits, and denote the restriction of $U_g$ to $\HH_{s,m}$ by
$W_m(g)$. The group contains $|G_s|=3^s s!$ elements. Three phases are
convenient because the same phase set will resolve the difference
patterns used in the separable comparison of Appendix~\ref{sec:resource}.

\begin{lemma}[Irreducible sector action]
\label{lem:irrep}
For each $s,m$, the representation $W_m$ is irreducible. Moreover,
\begin{equation}
 \AD^{\otimes s}(U_gTU_g^\dagger)
   =U_g\AD^{\otimes s}(T)U_g^\dagger.
 \label{eq:covariance}
\end{equation}
\end{lemma}
\begin{proof}
On the subset basis, $D_{\bm t}$ has character
$\omega_3^{\sum_{j\in S}t_j}$. Two different subsets have different
characters, since a phase at a site in their symmetric difference
distinguishes them. An operator commuting with all $D_{\bm t}$ is
therefore diagonal in this basis. Commutation with all permutations
forces all its diagonal entries to coincide: permutations act
transitively on subsets of size $m$. The commutant is scalar, proving
irreducibility, including the one-dimensional endpoint sectors.

For a single qubit, $\AD$ leaves diagonal operators diagonal and maps
$\ket0\bra1$ to $\sqrt\eta\ket0\bra1$. It consequently commutes with
phase rotations. Tensor products give covariance under independent
phases, and identical channel factors commute with permutations.
Together these facts prove Eq.~\eqref{eq:covariance}.
\end{proof}

\begin{proposition}[Qubit sector simulation]
\label{prop:simulation}
Suppose a channel $\mathcal N_s$ with $s$ input qubits satisfies
\begin{equation}
 \mathcal N_s(U_gTU_g^\dagger)
      =V_g\mathcal N_s(T)V_g^\dagger
 \label{eq:generalcov}
\end{equation}
for output unitaries $V_g$. Then its restriction to $\HH_{s,m}$ admits
an exact finite-outcome LOCC simulation with resource
\begin{equation}
 J^{\mathcal N}_{s,m}=(\id\otimes\mathcal N_s)(\Phi_{s,m}).
 \label{eq:resourcegeneral}
\end{equation}
Writing the corresponding LOCC as $\mathcal T^{\mathcal N}_{s,m}$,
the simulation identity is
\begin{equation}
 \mathcal N_s(\rho_A)
 =\mathcal T^{\mathcal N}_{s,m}
    \!\left(\rho_A\otimes J^{\mathcal N}_{s,m}\right),
 \label{eq:general-simulation-LOCC}
\end{equation}
for every input supported in $\HH_{s,m}$. The identity remains valid
after adjoining the identity channel on arbitrary external memories
held by either party.
\end{proposition}
\begin{proof}
Lemma~\ref{lem:irrep} verifies the input irreducibility required by the
covariant teleportation construction of Ref.~\cite{GBW},
Proposition~24, used in Ref.~\cite{Bosonic}, Appendix~B.1.
For completeness, its implementation here is explicit. Write $d=d_{s,m}$
and let Alice measure her input and her resource share with effects
\begin{equation}
 M_g=\frac{d^2}{|G_s|}\proj{\beta_g},~~
 \ket{\beta_g}=(W_m(g)\otimes I)\ket{\Phi_{s,m}}.
 \label{eq:Bell}
\end{equation}
Schur twirling gives $\sum_gM_g=I$ on the two sector spaces.
Let $L_g=M_g^{1/2}\otimes I_B$. The teleportation LOCC, acting on an
input $\xi_{ARB}$ with $A R$ held by Alice and $B$ by Bob, is
\begin{equation}
 \begin{split}
 \mathcal T^{\mathcal N}_{s,m}(\xi_{ARB})
 &=\sum_{g\in G_s}V_g\xi_B^gV_g^\dagger,\\
 \xi_B^g&=\Tr_{AR}(L_g\xi_{ARB}L_g^\dagger).
 \end{split}
 \label{eq:teleportation-LOCC-map}
\end{equation}
Alice's measurement, her classical message $g$, and Bob's correction
$V_g$ make this a deterministic, one-way \mbox{LOCC} channel on the sector
spaces. With these conventions the
unnormalized state before correction is
$|G_s|^{-1}\mathcal N_s(W_m(g)^\dagger\rho W_m(g))$;
Eq.~\eqref{eq:generalcov} makes the corrected branch
$|G_s|^{-1}\mathcal N_s(\rho)$. This is the cited teleportation identity,
with the qubit group substituted. Its complete, rather than merely
statewise, channel identity includes the external memories.
\end{proof}

For $\mathcal N_s=\AD^{\otimes s}$ take $V_g=U_g$ and write the resource
as $J_{s,m}$. Equation~\eqref{eq:teleportation-LOCC-map} then defines
the map $\mathcal T_{s,m}$ used in
Eqs.~\eqref{eq:sector-LOCC-main} and \eqref{eq:sector-LOCC-memory}.
Its measurement and correction depend on $s,m$, not on $\gamma$.
All correction operations preserve the qubit Hilbert space.
The proposition also applies to tensor powers of other phase-covariant
qubit channels, including generalized amplitude damping, dephasing,
and depolarization. Exact simulation alone does not establish a
sharper converse: a channel-dependent entanglement estimate is
also required. Appendix~\ref{sec:gad-proof} supplies such an estimate
for generalized amplitude damping.

Finally note that the exact simulation is a statement about an individual sector.
It does not assert that an arbitrary input may be dephased across
sectors without a cost. Appendix~\ref{sec:adaptive} accounts for those
coherences using the established pinching inequality.

\section{Resource geometry, separable optimality, and additivity}
\label{sec:resource}

For a bipartite state, the relative entropy of entanglement (REE) and its
regularization are defined from the quantum relative entropy
~\cite{Umegaki,Vedral,VP} by
\begin{align}
 \ER(\rho)&=\inf_{\sigma\in\SEP}D(\rho\|\sigma),\nonumber\\
 \ERinf(\rho)&=\lim_{r\to\infty}\frac{1}{r}\ER(\rho^{\otimes r}),
 \label{eq:REE}
\end{align}
where $D(\rho\|\sigma)=\Tr\rho(\log_2\rho-\log_2\sigma)$ with the usual
support convention, and $\SEP$ is the trace-norm closed convex hull of
product states~\cite{Vedral,VP}. We use LOCC monotonicity, convexity,
subadditivity, and the local-flags property~\cite{Vidal,PlenioVirmani}.
Finite-dimensional continuity is established in
Ref.~\cite{DonaldHorodecki}. For normal states on
separable spaces, lower semicontinuity supplies the extension used in
the bosonic analysis~\cite{Bosonic}.

\subsection{Conditional resource geometry}

The orthogonal output-number decomposition is Eq.~\eqref{eq:resourceblocks}.
For the branch with $k$ survivors, write $\ell=m-k$ for the number
of lost excitations.
For each loss set $L\subseteq[s]$ with $|L|=\ell$, define
\begin{equation}
 \ket{\psi_L}=\frac{1}{\sqrt{\binom{s-\ell}{k}}}
   \sum_{\substack{|T|=k\\T\cap L=\varnothing}}
          \ket{L\cup T}_R\ket T_B.
 \label{eq:psiL}
\end{equation}
Then
\begin{equation}
 \tau_{s,m,k}=\frac{P_{s,m,k}}{\binom{s}{\ell}},~~
 P_{s,m,k}=\sum_{|L|=\ell}\proj{\psi_L}.
 \label{eq:tau}
\end{equation}

To verify these formulas directly, apply the tensor-product Kraus
operator with $A_1$ on $L$ and $A_0$ on its complement to the second half
of Eq.~\eqref{eq:Phi}. It annihilates subsets not containing $L$; for the
others its amplitude is $\gamma^{\ell/2}\eta^{k/2}$, and it leaves the
surviving set $T=S\setminus L$. The squared norm of this branch is
$\gamma^\ell\eta^k\binom{s-\ell}{k}/\binom{s}{m}$.
The loss vectors are orthonormal, because every basis pair
$(L\cup T,T)$ determines $L$. Summing over loss sets uses
\begin{equation}
 \binom{s}{\ell}\binom{s-\ell}{k}
     =\binom{s}{m}\binom{m}{k},
 \label{eq:counting}
\end{equation}
which gives both the binomial probabilities and Eq.~\eqref{eq:tau}.
In particular, the conditional resource is maximally mixed on a support
of rank $\binom{s}{\ell}$.

\subsection{Separable comparison on the resource support}

Symmetry can reduce the optimization of entanglement measures
~\cite{VW}. Here we construct the comparison explicitly and
prove its global optimality below. Introduce normalized
phase-twisted Dicke vectors
\begin{equation}
 \ket{D_{s,r}(\bm t)}=\frac{1}{\sqrt{\binom{s}{r}}}
       \sum_{|S|=r}\omega_3^{\sum_{j\in S}t_j}\ket S,
 \label{eq:Dicke}
\end{equation}
and the separable state
\begin{equation}
 \sigma_{s,m,k}=\frac{1}{3^s}\sum_{\bm t\in\ZZ_3^s}
       D_{s,m}(\bm t)_R\otimes D_{s,k}(-\bm t)_B.
 \label{eq:sigma}
\end{equation}
Each Dicke vector can be entangled among qubits on one side; separability
here, as required, is across the bipartition $R:B$.

\begin{lemma}[Conditional REE upper bound]
\label{lem:comparison}
For $0\leq k\leq m\leq s$, let
\begin{equation}
 w_{s,m,k}=\frac{\binom{m}{k}}{\binom{s}{k}}.
 \label{eq:w}
\end{equation}
The state in Eq.~\eqref{eq:sigma} obeys
\begin{equation}
 \sigma_{s,m,k}P_{s,m,k}=w_{s,m,k}\tau_{s,m,k}.
 \label{eq:supportidentity}
\end{equation}
Consequently,
\begin{equation}
 \ER(\tau_{s,m,k})\leq
 D(\tau_{s,m,k}\|\sigma_{s,m,k})
   =\log_2\frac{\binom{s}{k}}{\binom{m}{k}}.
 \label{eq:conditionalbound}
\end{equation}
\end{lemma}
\begin{proof}
The amplitude of the product vector in Eq.~\eqref{eq:sigma} at a basis
pair $(S,T)$ is
\begin{equation}
 \frac{\omega_3^{\bm t\cdot(\bm1_S-\bm1_T)}}
      {\sqrt{\binom{s}{m}\binom{s}{k}}},
 \label{eq:phaseamplitude}
\end{equation}
where $\bm1_S$ is the indicator vector of $S$. The average over each
$t_j\in\ZZ_3$ removes a matrix element unless the corresponding
difference vectors agree modulo three. Their entries lie in
$\{-1,0,1\}$, so agreement modulo three is exact agreement. Hence
$\sigma_{s,m,k}$ is a direct sum of rank-one blocks indexed by
$\bm z=\bm1_S-\bm1_T$.

For $\bm z=\bm1_L$, the pairs in that block are precisely
$S=L\cup T$ with $T\cap L=\varnothing$, $|T|=k$. The normalized vector
of the block is $\ket{\psi_L}$, and its eigenvalue is
\begin{equation}
 \lambda_{s,m,k}
   =\frac{\binom{s-\ell}{k}}{\binom{s}{m}\binom{s}{k}}.
 \label{eq:lambda}
\end{equation}
All remaining difference patterns are orthogonal to these blocks.
By Eq.~\eqref{eq:counting},
$\binom{s}{\ell}\lambda_{s,m,k}=w_{s,m,k}$.
Using Eq.~\eqref{eq:tau} therefore proves
Eq.~\eqref{eq:supportidentity}, including the absence of matrix elements
coupling the resource support to its orthogonal complement.

On that support $\tau_{s,m,k}$ has eigenvalue
$1/\binom{s}{\ell}$ and $\sigma_{s,m,k}$ has eigenvalue
$\lambda_{s,m,k}$. Their logarithms differ by $-\log_2w_{s,m,k}$,
which proves the relative-entropy identity and the upper bound.
\end{proof}

The comparison supplies the upper bound without an optimization.
We next prove that it is optimal among all separable states.
For $k=0$ or $m=s$, its cost is zero; for $k=m$, its cost is
$\log_2\binom{s}{m}$, the entropy of the maximally entangled sector.

\subsection{Optimality of the separable comparison}
\label{sec:optimality}

The missing lower bound is a sharp estimate of the largest separable
overlap with $P_{s,m,k}$. We use the following partition form of the
discrete Brascamp--Lieb inequality~\cite{BCLM}, Sec.~4.2.2
(Proposition~20 in its arXiv version). If $(A,B,C)$ is a uniformly
distributed ordered partition of $[s]$ with prescribed sizes
$u,v,w$, then nonnegative functions satisfy
\begin{equation}
 \EE[F(A)G(B)H(C)]
 \leq\sqrt{\EE F(A)^2\,\EE G(B)^2\,\EE H(C)^2}.
 \label{eq:partitionBL}
\end{equation}

To apply the cited proposition, fix a partition $(I_1,I_2,I_3)$ of
$[s]$ with block sizes $u,v,w$. For a uniform random permutation
$\pi$, the image sets $(\pi(I_1),\pi(I_2),\pi(I_3))$ have the required
distribution of $(A,B,C)$. For functions of these image sets, the
hypothesis is that, for every pair of distinct sites $i,j$,
\begin{equation}
 \sum_{r:\,|I_r\cap\{i,j\}|=1} c_r \leq 1.
\end{equation}
Take $c_1=c_2=c_3=1/2$. If $i$ and $j$ lie in the same block, the
sum is zero. If they lie in different blocks, exactly two blocks
contribute, so the sum is one. Applying the proposition to
$F^2,G^2,H^2$ with these weights gives the displayed inequality.
An empty block contributes only a constant factor.


\begin{theorem}[Optimal conditional comparison]
\label{thm:optimal}
For every $0\leq k\leq m\leq s$,
\begin{equation}
 \max_{\sigma\in\SEP}\Tr(P_{s,m,k}\sigma)
       =w_{s,m,k}=\frac{\binom{m}{k}}{\binom{s}{k}},
 \label{eq:overlapexact}
\end{equation}
and the comparison in Eq.~\eqref{eq:sigma} is REE-optimal, giving
the value in Eq.~\eqref{eq:REEexact}. Its regularized equality
follows from the additivity argument below.
\end{theorem}
\begin{proof}
It suffices to optimize the overlap over normalized product vectors
$\ket a=\sum_{|S|=m}a_S\ket S$ and
$\ket b=\sum_{|T|=k}b_T\ket T$; components outside these local sectors
cannot increase the maximum. Put $\ell=m-k$ and
$q=\binom{s-\ell}{k}$. From Eq.~\eqref{eq:psiL},
\begin{equation}
 \bra{ab}P_{s,m,k}\ket{ab}
 =\frac1q\sum_{|L|=\ell}
    \left|\sum_{\substack{|T|=k\\T\cap L=\varnothing}}
                        a_{L\cup T}b_T\right|^2.
 \label{eq:overlapnorm}
\end{equation}
By Euclidean duality, the square root of the right side is
\begin{equation}
 \frac1{\sqrt q}\max_{\sum_L|c_L|^2=1}
 \left|\sum_{\substack{|L|=\ell,\ |T|=k\\L\cap T=\varnothing}}
                   a_{L\cup T}b_Tc_L\right|.
 \label{eq:overlapdual}
\end{equation}
Replace the coefficients by their absolute values for an upper bound.
Apply Eq.~\eqref{eq:partitionBL} to the partition
$(A,B,C)=([s]\setminus(L\cup T),T,L)$, with
\begin{equation}
 F(A)=|a_{[s]\setminus A}|,~~ G(B)=|b_B|,~~ H(C)=|c_C|.
 \label{eq:BLfunctions}
\end{equation}
There are $Z=\binom{s}{m}\binom{m}{k}=\binom{s}{\ell}q$ such
partitions. Their marginals are uniform on subsets of the specified
sizes, giving
\begin{equation}
 \left|\sum_{L,T\,\mathrm{disjoint}}a_{L\cup T}b_Tc_L\right|
 \leq\frac{Z}{\sqrt{\binom{s}{m}\binom{s}{k}\binom{s}{\ell}}}.
 \label{eq:BLapplied}
\end{equation}
Squaring and dividing by $q$ gives exactly $w_{s,m,k}$.
Constant, normalized coefficients $a_S$ and $b_T$ attain it.
Convexity extends the bound to every separable $\sigma$, proving
Eq.~\eqref{eq:overlapexact}.

Finally apply relative-entropy data processing~\cite{Lindblad} to the two-outcome
measurement $\{P_{s,m,k},I-P_{s,m,k}\}$. Since the resource passes with
probability one, every separable comparison satisfies
\begin{equation}
 D(\tau_{s,m,k}\|\sigma)
 \geq-\log_2\Tr(P_{s,m,k}\sigma)\geq-\log_2w_{s,m,k}.
 \label{eq:REElower}
\end{equation}
Lemma~\ref{lem:comparison} attains this lower bound.
\end{proof}

\subsection{Tensor products and exact unconditional resource cost}
\label{sec:additivity}

The optimality survives joint separable comparisons across resources.
This requires a proof: regularization and tensor-product
optimization are substantive issues for relative-entropy
entanglement bounds~\cite{VW,Audenaert,Rains2001}.
For a positive bipartite operator $T$, define
\begin{equation}
 \nu(T)=\max_{\|a\|=\|b\|=1}\bra{ab}T\ket{ab}.
 \label{eq:nu}
\end{equation}
The nonnegative-operator multiplicativity used in geometric
entanglement~\cite{ZCH} states that
\begin{equation}
 \nu(P\otimes Q)=\nu(P)\nu(Q)
 \label{eq:multiplicativity}
\end{equation}
if $P$ is positive with nonnegative entries in a product basis and
$Q$ is positive. The bipartition groups the two Alice factors and the
two Bob factors. To recall the relevant argument, write arbitrary local
vectors as $\sum_i\ket i\ket{a_i}$ and $\sum_j\ket j\ket{b_j}$.
Positivity of $Q$ gives
\begin{equation}
 |\bra{a_i b_j}Q\ket{a_k b_l}|
 \leq\nu(Q)\|a_i\|\|b_j\|\|a_k\|\|b_l\|.
 \label{eq:positiveCS}
\end{equation}
Multiplication by the nonnegative entries $P_{ij,kl}$ and summation
bound the joint overlap by $\nu(Q)\nu(P)$; product optimizers give the
reverse inequality. This argument allows arbitrary entanglement within
either party's collection of factors.

Every $P_{s,m,k}$ has nonnegative entries by
Eqs.~\eqref{eq:psiL}--\eqref{eq:tau}. Therefore, for any finite list
$\tau_j=\tau_{s_j,m_j,k_j}$,
\begin{equation}
 \nu\!\left(\bigotimes_jP_{s_j,m_j,k_j}\right)=\prod_jw_{s_j,m_j,k_j}.
 \label{eq:productoverlap}
\end{equation}
The support-test lower bound in Eq.~\eqref{eq:REElower} and the product
of the optimal comparisons then prove
\begin{equation}
 \begin{split}
 \ER\!\left(\bigotimes_j\tau_j\right)&=\sum_j\ER(\tau_j),\\
 \ERinf(\tau_{s,m,k})&=\ER(\tau_{s,m,k}).
 \end{split}
 \label{eq:conditionaladditivity}
\end{equation}

Bob can distinguish the output-number blocks in $J_{s,m}$ locally.
The exact conditional REEs consequently give the average
$F_s(m)$ in Eq.~\eqref{eq:F}.
An optimal separable comparison is
$\Sigma_{s,m}=\sum_kp_{m,k}\sigma_{s,m,k}$.
Applying the same local-flags identity to tensor products and using
Eq.~\eqref{eq:conditionaladditivity} gives
\begin{equation}
 \begin{split}
 \ER\!\left(\bigotimes_jJ_{s_j,m_j}\right)&=\sum_jF_{s_j}(m_j),\\
 \ERinf(J_{s,m})&=F_s(m).
 \end{split}
 \label{eq:resourceadditivity}
\end{equation}
Thus neither single-resource optimization nor collective regularization
of these resources lowers their REE cost.

\subsection{From the sector label to excitation density}

The next estimate is the qubit resource input to the adaptive framework.

\begin{lemma}[Exact classical cost and finite density bound]
\label{lem:density}
Let $S$ be a uniformly distributed $m$-subset of $[s]$, and obtain
$T\subseteq S$ by retaining each element independently with probability
$\eta$. For every $0\leq m\leq s$,
\begin{equation}
 F_s(m)=I(S:T)\leq s f_\gamma(m/s).
 \label{eq:density}
\end{equation}
\end{lemma}
\begin{proof}
Writing $K=|T|$, the binomial distribution satisfies the identity
\begin{equation}
 \EE\log_2\binom{m}{K}=m h_2(\eta)-H(K).
 \label{eq:binomialentropy}
\end{equation}
The output set is uniform conditional on $K$, while the losses are
independent conditional on $S$. Hence
\begin{align}
 H(T)&=H(K)+\EE\log_2\binom{s}{K},\nonumber\\
 H(T|S)&=m h_2(\eta).
 \label{eq:classicalentropies}
\end{align}
Equations~\eqref{eq:F} and \eqref{eq:binomialentropy} now give
$F_s(m)=H(T)-H(T|S)=I(S:T)$. Each output indicator has mean
$\eta m/s$, so entropy subadditivity yields
\begin{equation}
 H(T)\leq\sum_{j=1}^s H(\bm1_{\{j\in T\}})
           =s h_2(\eta m/s).
 \label{eq:binomialupper}
\end{equation}
Subtracting $m h_2(\eta)$ proves Eq.~\eqref{eq:density}.
No logarithmic correction is needed for the resource cost itself.
\end{proof}

The estimate has the exact leading term. For every sequence
$m_s/s\to x\in[0,1]$,
\begin{equation}
 \lim_{s\to\infty}\frac{F_s(m_s)}s=f_\gamma(x).
 \label{eq:densityexact}
\end{equation}
Indeed, $\log_2\binom{s}{k}$ differs from $s h_2(k/s)$ by at most
$\log_2(s+1)$, uniformly in $k$. Under $K\sim\Bin(m_s,\eta)$,
$K/s\to\eta x$ in probability. Equation~\eqref{eq:binomialentropy},
boundedness and continuity of $h_2$, and $H(K)\leq\log_2(s+1)$
then prove Eq.~\eqref{eq:densityexact}. In particular, optimality
strengthens the resource theorem but does not decrease the density
used by the capacity converse.

\section{Application of the adaptive converse framework}
\label{sec:adaptive}

\subsection{Operational setting and statement}

We consider memoryless uses of $\AD$ interspersed with adaptive local
operations and authenticated public two-way classical communication
(LOCC). The parties have no initial shared entanglement or secret key.
States and instruments are normal on separable Hilbert spaces. Each
finite protocol uses finitely many local rounds with finite or countable
outcomes; its round bound may grow with the number of channel uses.
Local memories have no dimension constraint. 
Eve receives the full complementary output of a Stinespring dilation
of every channel use, including any purification of the bath,
and all public messages. These are the operational conventions
of Refs.~\cite{Bosonic,Noisy}.

The excitation convention is Eq.~\eqref{eq:constraint}; the
populations are unconditional, including all unsuccessful branches.
Target dimensions are finite at each blocklength and rates count
every channel use, consistent with operational entanglement
distillation~\cite{BDSW,Rains1999}. The unrestricted qubit problem has $N=1$.
We use the protocol $\mathcal P_n$, intermediate states $\rho_i$, and
final retained state $\zeta_n=\rho_n$ defined in
Eq.~\eqref{eq:protocol-state-main}. In particular,
$\rho^{(i)}_{A_i}=(\rho_{i-1})_{A_i}$, so the input constraint here
is exactly the one used in the proof sketch. All entanglement measures
are evaluated across the current Alice--Bob partition; the transmitted
qubit moves from Alice's side to Bob's at each channel use.

\begin{theorem}[Two-way capacity converse]
\label{thm:main}
Under the operational assumptions above, the four two-way capacities
satisfy Eqs.~\eqref{eq:main} and \eqref{eq:BN}. The unrestricted bound
is Eq.~\eqref{eq:B}. At $\gamma=0$ the constrained bound is
$h_2(\min\{N,1/2\})$, and at $\gamma=1$ it is zero.
\end{theorem}

We now substitute the qubit estimates into the framework of
Refs.~\cite{Bosonic,Noisy}. Only the cost of controlling coherence
between output-excitation sectors remains; it is logarithmic in the
block size.

\subsection{Excitation optimization}

For $0<\gamma<1$, direct differentiation gives
\begin{align}
 f_\gamma'(x)&=\eta\log_2\frac{1-\eta x}{\eta x}-h_2(\gamma),\nonumber\\
 f_\gamma''(x)&=-\frac{\eta}{(\ln2)x(1-\eta x)}<0.
 \label{eq:concavity}
\end{align}
Thus $f_\gamma$ is strictly concave, with
$f_\gamma(0)=f_\gamma(1)=0$, and its unique maximizer is
Eq.~\eqref{eq:xstar}. Substitution gives Eq.~\eqref{eq:B}.
For an upper constraint on $x$, the maximum occurs at
$\min\{N,x_\gamma^*\}$, explaining the plateau in
Eq.~\eqref{eq:BN}.

At $\gamma=0$, $f_0(x)=h_2(x)$; at $\gamma=1$, $f_1(x)=0$.
These yield the endpoint values in Theorem~\ref{thm:main}.

\subsection{Block entanglement gain}

Consider any input $\rho_{XA^sY}$ of finite
$\ER(XA^s:Y)_\rho$, where $X$ is Alice's memory and $Y$ is Bob's.
Let $x=\Tr(\widehat M_s\rho)/s$ and
$\omega=(\id_{XY}\otimes\AD^{\otimes s})(\rho)$. We obtain
\begin{equation}
 \begin{split}
 \ER(X:B^sY)_\omega\leq {}&\ER(XA^s:Y)_\rho\\
                  &+s f_\gamma(x)+\log_2(s+1).
 \end{split}
 \label{eq:block}
\end{equation}

Here is the channel-specific substitution. Let $\PP_A$ and $\PP_B$
pinch total input and output excitation, respectively. Continuous
phase covariance gives
\begin{equation}
 \PP_B\circ\AD^{\otimes s}
     =\AD^{\otimes s}\circ\PP_A.
 \label{eq:pinchcommute}
\end{equation}
If $p_m$ and $\rho_m$ are the input-sector probabilities and normalized
conditional states, exact simulation and LOCC monotonicity imply
\begin{equation}
 \ER(\PP_B\omega)
   \leq\sum_m p_m[\ER(\rho_m)+F_s(m)].
 \label{eq:pinchedcost}
\end{equation}
The first average is at most $\ER(\rho)$ by selective monotonicity.
Lemma~\ref{lem:density} and concavity in Eq.~\eqref{eq:concavity}
bound the second by $s f_\gamma(x)$.

Let $K$ denote the total output excitation number. The local-pinching
inequality of Ref.~\cite[Lemma~2]{Bosonic} gives
\begin{align}
 \ER(\omega)
 &\le \ER(\PP_B\omega)+H(K) \notag \\
 &\le \ER(\PP_B\omega)+\log_2(s+1),
\end{align}
since $K$ takes values in $\{0,\ldots,s\}$. This proves
Eq.~\eqref{eq:block} for finite-dimensional memories.

For arbitrary separable memory Hilbert spaces, apply the local
finite-rank compressions of Ref.~\cite[Appendix~B.4]{Bosonic}.
These trace-preserving maps act only on Alice's and Bob's memories,
so they preserve the joint marginal of the transmitted qubits and
hence the mean excitation $x$. By LOCC monotonicity, the REE of each compressed input is at most that of the original input. The
corresponding outputs converge in trace norm to $\omega$ as the
compression ranks increase. Lower semicontinuity of the REE therefore
gives Eq.~\eqref{eq:block} for the original state. The protocol inputs
may retain arbitrary coherence between excitation sectors.

\subsection{Replicas and operational targets}

Fix an $n$-use adaptive protocol $\mathcal P_n$ with retained final
state $\zeta_n$, as defined in Eq.~\eqref{eq:protocol-state-main}.
Following Ref.~\cite[Appendix~B.5]{Bosonic} and
Ref.~\cite[Appendix~C.2]{Noisy}, run $r$ independent copies of the
entire protocol. At round $i$, collect the $r$ transmitted qubits into
a single block. Each qubit has the same unconditional excitation
probability $x_i$, so the mean excitation per qubit in that block is
$x_i$. Independence is required only between replicas; successive
inputs within one adaptive run may be correlated.

The initial state is separable across Alice and Bob, and the LOCC
between transmissions cannot increase the REE. Applying
Eq.~\eqref{eq:block} to each of the $n$ replicated rounds therefore
gives
\begin{equation}
 \ER(\zeta_n^{\otimes r})
 \leq r\sum_{i=1}^n f_\gamma(x_i)+n\log_2(r+1).
 \label{eq:replica}
\end{equation}
Dividing by $r$ and taking $r\to\infty$ at fixed $n$ removes the
logarithmic term and yields
\begin{equation}
 \ERinf(\zeta_n)
 \leq \sum_{i=1}^n f_\gamma(x_i)
 \leq n\Bsec(\gamma,N).
 \label{eq:adaptive}
\end{equation}
The last inequality follows from the concavity of $f_\gamma$ and the
excitation constraint in Eq.~\eqref{eq:constraint}.
Equations~\eqref{eq:replica} and~\eqref{eq:adaptive} are the bounds
stated in Eqs.~\eqref{eq:replica-main} and~\eqref{eq:adaptive-main}
of the proof sketch.

To relate Eq.~\eqref{eq:adaptive} to a communication rate, let
$d=d_n$ be the target dimension at blocklength $n$. For entanglement
distribution, the test projects the output registers onto the target
maximally entangled state. For key generation, the privacy test acts
on the retained key registers and their local shields before the final
key measurement~\cite{PrivateParadigm}. The distinction matters:
the measured classical key alone is separable, whereas a separable
state across Alice and Bob passes a size-$d$ privacy test with
probability at most $1/d$. If a retained state $\zeta$ passes a
size-$d$ entanglement or privacy test with probability $a$, the
regularized test inequality of Ref.~\cite[Lemmas~8 and~11]{Bosonic}
and Ref.~\cite[Appendix~A.1]{Noisy} gives
\begin{equation}
 \ERinf(\zeta)\geq a\log_2 d-h_2(a).
 \label{eq:test}
\end{equation}

For a key protocol, the input-preserving reduction of
Ref.~\cite[Lemma~12]{Bosonic} yields a retained state $\zeta_n$ with
privacy-test acceptance $a\geq(1-e)^2$. The reduction preserves every
transmitted population $x_i$, so Eq.~\eqref{eq:adaptive} applies to
this same state with the original excitation budget. Here the key
error is
\begin{equation}
 e=\tfrac12\left\|\kappa_{K_AK_BE}
       -\tau_d^{\rm key}\otimes\kappa_E\right\|_1,
 \label{eq:keyerror}
\end{equation}
where $\kappa_{K_AK_BE}$ is the state after the key measurement,
$\tau_d^{\rm key}=d^{-1}\sum_{j=0}^{d-1}\proj{jj}$ is the ideal
shared key, and Eve's register $E$ includes the public transcript.

\begin{corollary}[Finite-error bounds]
\label{cor:finite}
For a target dimension $d$, an $n$-use entanglement-distribution
protocol with entanglement infidelity at most $\varepsilon$ satisfies
\begin{equation}
 (1-\varepsilon)\log_2 d
 \leq \sum_{i=1}^n f_\gamma(x_i)+1.
 \label{eq:finiteent}
\end{equation}
An $n$-use key protocol with error $e$ as defined in
Eq.~\eqref{eq:keyerror} satisfies
\begin{equation}
 (1-e)^2\log_2 d
 \leq \sum_{i=1}^n f_\gamma(x_i)+1.
 \label{eq:finitekey}
\end{equation}
In both cases, the sum is at most $n\Bsec(\gamma,N)$.
\end{corollary}

\begin{proof}
Apply Eq.~\eqref{eq:test} to the retained output $\zeta_n$ and combine
it with Eq.~\eqref{eq:adaptive}. Since $h_2(a)\leq 1$, this gives
\[
 a\log_2 d \leq \sum_{i=1}^n f_\gamma(x_i)+1.
\]
For entanglement distribution, the target projection succeeds with
probability $a\geq 1-\varepsilon$. For key generation, the retained
state passes the privacy test with probability $a\geq(1-e)^2$.
Substitution gives Eqs.~\eqref{eq:finiteent} and~\eqref{eq:finitekey},
respectively.
\end{proof}

Dividing the finite-error bounds by $n$ and taking $n\to\infty$ with
vanishing error bounds the achievable rate by $\Bsec(\gamma,N)$.
The standard equivalences between quantum communication and
entanglement distribution, and between private communication and
secret-key agreement, then give Theorem~\ref{thm:main}. These are
weak-converse bounds.

\section{Regularized output entanglement and sharp entanglement gain}
\label{sec:stateREE}

We prove Eq.~\eqref{eq:OmegaxREE} for the input and output states
defined in Eq.~\eqref{eq:Omegax}, and establish the sharpness of the
regularized-REE gain used by the converse.

\begin{theorem}[Exact regularized output entanglement]
\label{thm:regularized}
For every $0\leq\gamma,x\leq1$, the exact identity
$\ERinf(\Omega_{\gamma,x})=f_\gamma(x)$ holds.
Moreover, $f_\gamma(x)$ is the optimal universal bound on the increase
of regularized REE at fixed input excitation $x$, for finite-REE inputs
with arbitrary memories.
\end{theorem}
\begin{proof}
Pinch the total excitation of the reference in
$\Omega_{\gamma,x}^{\otimes s}$. The binomial Schmidt weights give
\begin{equation}
 \PP_R(\Omega_{\gamma,x}^{\otimes s})
  =\bigoplus_{m=0}^s\binom{s}{m}x^m(1-x)^{s-m}J_{s,m}.
 \label{eq:Omegapinched}
\end{equation}
The direct sum is locally flagged on $R$. Consequently, if
$M\sim\Bin(s,x)$, local monotonicity and the exact flags identity give
the first inequality in
\begin{equation}
 \EE_M F_s(M)\leq\ER(\Omega_{\gamma,x}^{\otimes s})
           \leq s f_\gamma(x)+\log_2(s+1).
 \label{eq:Omegasandwich}
\end{equation}
The second follows from Eq.~\eqref{eq:block}, with all of
$\varphi_x^{\otimes s}$ initially held by Alice. 
To evaluate the lower term, take $K|M\sim\Bin(M,\eta)$.
Binomial thinning gives $K\sim\Bin(s,\eta x)$.
Equations~\eqref{eq:F} and \eqref{eq:binomialentropy}
therefore imply the exact finite-block identity
\begin{equation}
 \begin{split}
 \EE_M F_s(M)
 &=\EE_K\log_2\binom{s}{K}-sx h_2(\eta)\\
 &~~+H(K|M)\\
 &=s f_\gamma(x)-I(M:K).
 \end{split}
 \label{eq:averageddensity}
\end{equation}
The second equality uses
\begin{equation}
 H(K)=s h_2(\eta x)-\EE_K\log_2\binom{s}{K}
\end{equation}
and $h_2(\eta)=h_2(\gamma)$.
Thus the difference between the averaged sector cost and
$s f_\gamma(x)$ is exactly the classical mutual information
between the input and surviving excitation counts.
Since
\begin{equation}
 0\leq I(M:K)\leq H(M)\leq\log_2(s+1),
\end{equation}
this difference is sublinear in $s$.
Combining Eq.~\eqref{eq:averageddensity} with
Eq.~\eqref{eq:Omegasandwich}, dividing by $s$, and taking
$s\to\infty$ gives
$\ERinf(\Omega_{\gamma,x})=f_\gamma(x)$,
proving Eq.~\eqref{eq:OmegaxREE}.

For an arbitrary finite-REE input $\rho_{XAY}$, let
$\omega=(\id_{XY}\otimes\AD)(\rho)$ and
$x=\Tr(\proj1\rho_A)$. Applying Eq.~\eqref{eq:block} to $r$ copies
and dividing by $r$ gives
\begin{equation}
 \ERinf(X:BY)_\omega-\ERinf(XA:Y)_\rho\leq f_\gamma(x).
 \label{eq:amortizedsharp}
\end{equation}
The input $\varphi_x$, prepared entirely by Alice with a trivial Bob
memory, has initial entanglement zero across the parties and attains
the right side by Eq.~\eqref{eq:OmegaxREE}. Hence the bound is optimal.
\end{proof}


Equation~\eqref{eq:resourceadditivity} gives the exact REE of the
sector resources, and Eq.~\eqref{eq:OmegaxREE} shows that the probe
attains the regularized entanglement gain used in the converse. Optimizing the separable comparison states therefore cannot lower
the bound in Eq.~\eqref{eq:B} within this framework. Whether a
communication protocol attains that bound remains open in general.

\subsection{When regularization lowers the probe entanglement}
\label{app:probe-strict}

\begin{corollary}[Single-copy versus regularized REE]\label{cor:probe-strict}
For $0<\gamma,x<1$,
\begin{equation}
E_R(\Omega_{\gamma,x})=E_R^\infty(\Omega_{\gamma,x})
 \ \Longleftrightarrow\ x=x_{\mathrm c}(\gamma)
 :=\frac{1}{2-\gamma}.
\end{equation}
At every other interior population the single-copy value is strictly
larger. Equality also holds on the boundary: the probe output is pure
for $\gamma=0$ and separable for $\gamma=1$ or $x\in\{0,1\}$.
\end{corollary}

\textit{Proof.}
First we record what equality would require of a closest separable
state. Let $\rho$ be a bipartite state on two qubits, $\sigma$ any
separable comparison, and $p$ and $p'$ their respective excitation
probabilities on one chosen side. With relative entropies in bits,
\begin{equation}
 D(\rho\Vert\sigma)\ \ge\
 E_R^\infty(\rho)+D_{\mathrm{bin}}(p\Vert p'),
 \label{eq:probe-marginal-penalty}
\end{equation}
where $D_{\mathrm{bin}}$ is the relative entropy of two Bernoulli laws.
Indeed, pinch the total excitation on that side of $n$ copies, denoting
the pinching by $\mathcal P$. The two count laws are binomial and
their relative entropy is $nD_{\mathrm{bin}}(p\Vert p')$, since the
log-likelihood ratio depends only on the count. Conditional blocks
of $\mathcal P(\sigma^{\otimes n})$ are separable. Data processing,
the relative-entropy chain rule for orthogonal blocks, and the local
flags identity therefore give
\begin{align}
 nD(\rho\Vert\sigma)\ \ge{}&
 nD_{\mathrm{bin}}(p\Vert p')
 + E_R\bigl(\mathcal P(\rho^{\otimes n})\bigr).
\end{align}
Local pinching loses at most $\log_2(n+1)$ in REE, as in Appendix~D.
Division by $n$ and passage to the limit give
Eq.~\eqref{eq:probe-marginal-penalty}. Applying it separately to
either qubit shows that if $E_R(\rho)=E_R^\infty(\rho)$, any optimal
one-copy comparison must match both local excitation probabilities.

Now put $\rho=\Omega_{\gamma,x}$ and write
\begin{gather}
 a=1-x,~~ b=\gamma x,~~ d=(1-\gamma)x,\\
 q=a+d,~~ u=1-a,~~ y=1-d,~~
 \alpha=a/q,~~ \delta=d/q.\nonumber
\end{gather}
The state has two orthogonal components,
\begin{align}
 |v\rangle&=\sqrt{\alpha}|00\rangle+\sqrt{\delta}|11\rangle,\\
 \rho&=q|v\rangle\langle v|+b|10\rangle\langle10|.
\end{align}
We keep $D$, $\ER$, and $\ERinf$ in bits. For the following
calculation, multiplication by $\ln 2$ converts the regularized
value to nats:
\begin{equation}
 (\ln 2)f_\gamma(x)=b\ln b-u\ln u-y\ln y.
\end{equation}

Suppose a separable $\sigma$ attains this value. A minimizer exists in
finite dimensions. Twirling it by $U_\theta\otimes U_{-\theta}$,
where $U_\theta=\operatorname{diag}(1,e^{i\theta})$, preserves
separability and optimality because $\rho$ is invariant. Write $S$
for the resulting $00,11$ block and $c$ for its $10$ population.
Finiteness implies $c>0$ and $|v\rangle\in\operatorname{supp} S$.
Set $L=\langle v|\ln S|v\rangle$ on this support.

Keep $v$ fixed and vary its weight from $q$ to $t\in(0,1)$. The states
$\rho_t=t|v\rangle\langle v|+(1-t)|10\rangle\langle10|$ are still
probes: $x_t=1-t\alpha$ and $\gamma_t=(1-t)/x_t$. The comparison
$\sigma$ is fixed. By the preceding theorem, its relative entropy
from $\rho_t$ is at least the regularized value for every $t$.
Subtracting that value in nats gives
\begin{align}
 G(t)={}&t\ln t-tL-(1-t)\ln c \\
 &+(1-t\alpha)\ln(1-t\alpha)
   +(1-t\delta)\ln(1-t\delta)\ \ge 0 .\nonumber
\end{align}
At the interior point $q$ we have $G(q)=G'(q)=0$. In particular
$G(q)-qG'(q)=\ln(uy/c)$, and then $G'(q)=0$ gives
\begin{equation}
 c=uy,~~ L=\ln q+\alpha\ln y+\delta\ln u.
\end{equation}
Since $\sigma$ matches both local excitation probabilities, its diagonal
in the order $00,01,10,11$ must now be $(ay,ad,uy,ud)$.

On the even block define $T=q\,\operatorname{diag}(y,u)$. The last
identity for $L$ reads $L=\langle v|\ln T|v\rangle$. On the other hand,
$T^{-1/2}ST^{-1/2}$ is positive semidefinite and has trace
$a/q+d/q=1$. If $S$ has rank two, this matrix has largest eigenvalue
$\lambda<1$. Thus $S\le\lambda T$. Operator monotonicity of the
logarithm gives
$L\le\ln\lambda+\langle v|\ln T|v\rangle$, a contradiction. If $S$
has rank one, its range must be spanned by $v$. Its diagonal entries
then have ratio $a/d$, whereas their prescribed ratio is $ay/(ud)$.
Hence $y=u$, or $a=d$. This is precisely $x=x_{\mathrm c}(\gamma)$.

Conversely, at $a=d$ write $a=d=z$ and $u=1-z$. The state is a
mixture of $\Phi^+=|\Phi^+\rangle\langle\Phi^+|$, with
$|\Phi^+\rangle=(|00\rangle+|11\rangle)/\sqrt 2$, and
$|10\rangle\langle10|$. The comparison
\begin{equation}
 \sigma_{\mathrm c}=2zu\,\Phi^+
       +z^2|01\rangle\langle01|+u^2|10\rangle\langle10|
\end{equation}
is separable: it has positive partial transpose in dimension
$2\times2$. Its eigenvalues on the occupied directions are $2zu$
and $u^2$. Direct substitution gives
$(\ln2)D(\rho\Vert\sigma_{\mathrm c})=b\ln b-2u\ln u
=(\ln2)f_\gamma(x)$. The preceding theorem gives the matching lower
bound. This proves the corollary.

For an independently quantified example, at $\gamma=1/2,x=1/3$ the
probe is exactly the state $\rho_{AB}(2/3,1/6)$ in Eq.~(14) of
Ref.~\cite{GalvaoPlenioVirmani}. Its single-copy REE lies within
$3.1\times10^{-8}$ bits of $0.354761489848$ [their Eq.~(25)],
whereas $E_R^\infty=h_2(1/6)-1/3=0.3166890883\ldots$ bits.
The strict inequality concerns regularization of the probe state;
it does not contradict additivity of the fixed-sector resources
or specify the first tensor power at which strict subadditivity occurs.


\section{Analytic comparison and high-damping limit}
\label{sec:comparisonproof}

The PLOB expression in Eq.~\eqref{eq:PLOB} is Supplementary
Eq.~(S211) of Ref.~\cite{PLOB}. Equation~\eqref{eq:sq} retains the input
maximization in its Supplementary Eq.~(S231); analytical evaluations
are also considered in Ref.~\cite{ADseries}. We compare the fully optimized balanced
construction with the sector bound.

\begin{proposition}[Strict analytic improvement]
\label{prop:improvement}
For every $0<\gamma<1$,
the strict inequalities in Eq.~\eqref{eq:strict} hold.
\end{proposition}
\begin{proof}
Since $0<t_\gamma<1$, Eq.~\eqref{eq:B} gives
$\Bsec<\log_2(1+\eta)=\Umax$ and $\Bsec<1$. Moreover,
\begin{equation}
 \gamma(1+\eta t_\gamma)<1
       ~~\Longleftrightarrow~~
 \gamma t_\gamma=\gamma^{1/\eta}<1,
 \label{eq:PLOBcomparison}
\end{equation}
so $\Bsec<-\log_2\gamma$. For the squashed bound, put $a=1-\gamma/2$, $b=\gamma/2$, so
$\eta=a-b$, and set $d_\gamma(x)=s_\gamma(x)-f_\gamma(x)$.
Its endpoints vanish. Direct differentiation gives
\begin{equation}
 d_\gamma''(x)=
 -\frac{b\eta(2-ax)}{(\ln2)(1-ax)(1-bx)(1-\eta x)}<0.
 \label{eq:gapsecond}
\end{equation}
Thus $d_\gamma(x)>0$ for $0<x<1$. Since $x_\gamma^*$ is interior,
\begin{equation}
 \Bsec=f_\gamma(x_\gamma^*)
      <s_\gamma(x_\gamma^*)\leq\Usq,
 \label{eq:sqcomparison}
\end{equation}
completing the proof.
\end{proof}

For completeness, the high-damping expansion uses $\eta=1-\gamma$:
\begin{equation}
 \begin{split}
 \log t_{1-\eta}
 &=\frac{1-\eta}{\eta}\log(1-\eta)\\
 &=-1+\frac{\eta}{2}+O(\eta^2),
 \end{split}
 \label{eq:t-expansion}
\end{equation}
where the logarithms in this expansion are natural. Therefore
$t_{1-\eta}=e^{-1}[1+O(\eta)]$, and expanding
$\log_2(1+\eta t_{1-\eta})$ gives Eq.~\eqref{eq:asymptotic}.
Dividing the exact triple-rail rate by that leading term proves
Eq.~\eqref{eq:ratio}.

\section{Generalized amplitude damping: simulation and adaptive converse}
\label{sec:gad-proof}

This appendix derives Eqs.~\eqref{eq:gad-capacities}--\eqref{eq:gad-energy}
from the amplitude-damping gain in Eq.~\eqref{eq:amortizedsharp}.
We reuse the operational setting and retained-state conversion in
Appendix~\ref{sec:adaptive}. No additional coherence assumption on
adaptive channel inputs is imposed.

\subsection{Channel action and exact sector simulation}

A Kraus representation of Eq.~\eqref{eq:gad-channel} is
\begin{align}
 K_0&=\sqrt{1-q}(\proj0+\sqrt\eta\proj1),\nonumber\\
 K_1&=\sqrt{(1-q)\gamma}\ket0\bra1,\nonumber\\
 K_2&=\sqrt q(\sqrt\eta\proj0+\proj1),\nonumber\\
 K_3&=\sqrt{q\gamma}\ket1\bra0.
 \label{eq:gad-kraus}
\end{align}
Thus, for $\rho=\left(\begin{smallmatrix}1-x&c\\c^*&x\end{smallmatrix}\right)$,
\begin{equation}
 \GAD(\rho)=
 \begin{pmatrix}
 1-\eta x-\gamma q&\sqrt\eta c\\
 \sqrt\eta c^*&\eta x+\gamma q
 \end{pmatrix}.
 \label{eq:gad-matrix}
\end{equation}
This action commutes with every local phase rotation. Taking tensor
powers and using identical factors gives, for the group of
Eq.~\eqref{eq:group},
\begin{equation}
 \GAD^{\otimes s}(U_gTU_g^\dagger)
 =U_g\GAD^{\otimes s}(T)U_g^\dagger.
 \label{eq:gad-covariance}
\end{equation}
Lemma~\ref{lem:irrep} already establishes the irreducible action on
$\HH_{s,m}$. Proposition~\ref{prop:simulation}, with $V_g=U_g$,
therefore proves Eq.~\eqref{eq:gad-sector-LOCC} with resource
$J_{s,m}^{\gamma,q}$ and the same map $\mathcal T_{s,m}$ as for
amplitude damping. Its measurement and corrections are independent
of $\gamma,q$.

\subsection{Convexity of the regularized entanglement gain}

For a channel $\mathcal N:A\to B$ and a state $\rho_{XAY}$ with
finite initial REE, define
\begin{equation}
 \Delta_{\mathcal N}(\rho)=
 \ERinf(X:BY)_{(\id_{XY}\otimes\mathcal N)(\rho)}
 -\ERinf(XA:Y)_\rho.
 \label{eq:gad-gain-definition}
\end{equation}
Here $X$ and $Y$ are arbitrary local memories. We use the following
consequence of the usual REE properties.

\begin{lemma}[Mixtures of entanglement gains]
\label{lem:gad-convexity}
Let $\mathcal N=\sum_j p_j\mathcal N_j$ be a finite channel mixture.
Suppose that the input $\rho_{XAY}$ and each branch output
$(\id_{XY}\otimes\mathcal N_j)(\rho)$ with $p_j>0$ have finite
single-copy REE across their respective Alice-Bob cuts. Then
\begin{equation}
 \Delta_{\mathcal N}(\rho)
 \leq\sum_jp_j\Delta_{\mathcal N_j}(\rho).
 \label{eq:gad-gain-convexity}
\end{equation}
\end{lemma}
\begin{proof}
We spell out why regularization preserves the convexity needed
here. Let $\omega=\sum_jp_j\omega_j$ with finite-REE states
$\omega_j$. Expanding $\omega^{\otimes n}$ and applying convexity
and subadditivity of single-copy REE gives
\begin{equation}
 \ER(\omega^{\otimes n})
 \leq\EE\sum_j\ER(\omega_j^{\otimes M_j}),
 \label{eq:gad-type-convexity}
\end{equation}
where $M_j$ is the number of occurrences of branch $j$ and
$\EE M_j=np_j$. For every $\delta>0$, convergence of the
regularized sequence supplies a finite $c_j$ such that
\begin{equation}
 \ER(\omega_j^{\otimes m})
 \leq m[\ERinf(\omega_j)+\delta]+c_j
 \label{eq:gad-regularization-estimate}
\end{equation}
for all integers $m\geq0$. Dividing Eq.~\eqref{eq:gad-type-convexity}
by $n$, taking $n\to\infty$, and then $\delta\downarrow0$ proves
$\ERinf(\omega)\leq\sum_jp_j\ERinf(\omega_j)$.
Apply this to the channel outputs. Their initial entanglement
term is the same, proving Eq.~\eqref{eq:gad-gain-convexity}.
\end{proof}

This lemma concerns the output entanglement measure for a fixed
input. It makes no assumption that any operational capacity is
convex in the channel. The finite-REE condition is preserved during
the protocols considered here: sending one qubit can increase
single-copy REE by at most one, and intervening LOCC cannot increase it.

\subsection{Three component estimates}

First, Eq.~\eqref{eq:amortizedsharp} gives, with
$x=\Tr(\proj1\rho_A)$,
\begin{equation}
 \Delta_{\AD}(\rho)\leq f_\gamma(x),~~
 \Delta_{\AD^\uparrow}(\rho)\leq f_\gamma(1-x).
 \label{eq:gad-ad-components}
\end{equation}
The second inequality follows by pre- and postcomposing with local
bit flips. Lemma~\ref{lem:gad-convexity} applied to
Eq.~\eqref{eq:gad-channel} therefore yields
\begin{equation}
 \Delta_{\GAD}(\rho)\leq F_{\gamma,q}(x).
 \label{eq:gad-direct-gain}
\end{equation}

Second, $\mathcal G_{\gamma,1/2}$ has Bloch contraction factors
$(\sqrt\eta,\sqrt\eta,\eta)$ and is Pauli covariant. Its normalized
Choi state is Bell diagonal, with probabilities
\begin{equation}
 \left\{\frac{(1+\sqrt\eta)^2}{4},
       \frac\gamma4,\frac\gamma4,
       \frac{(1-\sqrt\eta)^2}{4}\right\}.
 \label{eq:gad-bell-spectrum}
\end{equation}
The known Bell-diagonal REE formula gives $P_\gamma$ in
Eq.~\eqref{eq:gad-pauli}~\cite{VP,PLOB,KSW}. Exact Choi-state
simulation, LOCC monotonicity, and subadditivity of regularized
REE imply
\begin{equation}
 \Delta_{\mathcal G_{\gamma,1/2}}(\rho)\leq P_\gamma.
 \label{eq:gad-unital-gain}
\end{equation}
Only an upper estimate on the regularized Choi REE is needed in
this step: $\ERinf(J)\leq\ER(J)=P_\gamma$ suffices.

Third, every entanglement-breaking channel can be implemented by
measurement, a classical message, and state preparation at the
receiver. It consequently has nonpositive gain:
\begin{equation}
 \Delta_{\mathcal N_{\rm EB}}(\rho)\leq0.
 \label{eq:gad-eb-gain}
\end{equation}
The partial transpose of the qubit Choi state gives the known
condition \eqref{eq:gad-eb}~\cite{KSW}. 
For $\gamma_c\leq\gamma<1$, the smallest bath population satisfying
Eq.~\eqref{eq:gad-eb} is $q=r_c(\gamma)$. At $\gamma=1$, the channel
is a replacement channel for every bath population.

\subsection{Mixture profiles and the unrestricted expression}

Assume first $q\leq1/2$. Combining
Eqs.~\eqref{eq:gad-unitalmix-main}, \eqref{eq:gad-gain-convexity},
\eqref{eq:gad-ad-components}, and \eqref{eq:gad-unital-gain} gives
\begin{equation}
 \Delta_{\GAD}(\rho)
 \leq(1-2q)f_\gamma(x)+2qP_\gamma.
 \label{eq:gad-first-profile}
\end{equation}
For $\gamma_c\leq\gamma<1$ and $q<r_c(\gamma)$, affine dependence
on the bath population gives the stronger decomposition
\begin{align}
 \GAD={}&\left(1-\frac q{r_c(\gamma)}\right)\AD\nonumber\\
       &+\frac q{r_c(\gamma)}\mathcal G_{\gamma,r_c(\gamma)}.
 \label{eq:gad-eb-decomposition}
\end{align}
Equations~\eqref{eq:gad-gain-convexity} and
\eqref{eq:gad-eb-gain} then imply
\begin{equation}
 \Delta_{\GAD}(\rho)
 \leq\left(1-\frac q{r_c(\gamma)}\right)f_\gamma(x).
 \label{eq:gad-second-profile}
\end{equation}
For $r_c(\gamma)\leq q\leq1/2$, the original channel is itself
entanglement breaking. At $\gamma=1$ it is a replacement channel.
The identity
\begin{equation}
 \mathcal G_{\gamma,q}(\rho)
 =X\mathcal G_{\gamma,1-q}(X\rho X)X
 \label{eq:gad-reflection}
\end{equation}
extends these estimates to $q>1/2$, with input population $1-x$.
Together they prove
\begin{equation}
 \Delta_{\GAD}(\rho)\leq H_{\gamma,q}(x).
 \label{eq:gad-h-gain}
\end{equation}
Maximizing over $x\in[0,1]$ and using
$\max_x f_\gamma(x)=\Bsec(\gamma)$ gives
Eq.~\eqref{eq:gad-mix}. At $\gamma=\gamma_c$, the branches agree
because $r_c=1/2$ and $P_{\gamma_c}=0$.

\subsection{Adaptive protocols and the mean-excitation constraint}

Use the same protocol notation as in
Eq.~\eqref{eq:protocol-state-main}, with $\AD$ replaced by $\GAD$.
Thus $\zeta_n=\rho_n$ is again the retained final state, and
$x_i=\Tr[\proj1(\rho_{i-1})_{A_i}]$ is the unconditional population
sent on use $i$. The input-preserving retained-state construction of
Appendix~\ref{sec:adaptive} applies to these protocols as well.

Both $F_{\gamma,q}$ and $H_{\gamma,q}$ are concave and nonnegative
on $[0,1]$. Their pointwise minimum
\begin{equation}
 u_{\gamma,q}(x)=\min\{F_{\gamma,q}(x),H_{\gamma,q}(x)\}
 \label{eq:gad-combined-profile}
\end{equation}
is also concave: its hypograph is the intersection of the two
convex hypographs. Equations~\eqref{eq:gad-direct-gain} and
\eqref{eq:gad-h-gain} bound the regularized-REE increment on each
adaptive use, including its local memories, by $u_{\gamma,q}(x_i)$.
LOCC monotonicity between uses and telescoping from an initially
separable state give
\begin{equation}
 \ERinf(\zeta_n)\leq\sum_{i=1}^n u_{\gamma,q}(x_i)
 \leq n\max_{0\leq x\leq N}u_{\gamma,q}(x).
 \label{eq:gad-adaptive}
\end{equation}
The final step is Jensen's inequality and the unconditional mean
constraint. The retained-state test inequality \eqref{eq:test}
now gives the same finite-error factors as
Eqs.~\eqref{eq:finiteent} and \eqref{eq:finitekey}, with
$\sum_i f_\gamma(x_i)$ replaced by
$\sum_i u_{\gamma,q}(x_i)$. Taking vanishing error proves
Eq.~\eqref{eq:gad-energy}; bounding each increment by
$\Umix(\gamma,q)$ proves Eq.~\eqref{eq:gad-capacities}.
These are weak converses. No new strong-converse or exact-capacity
claim follows from this extension.

As a resource-level consequence, prepare $\Phi_{s,m}$ entirely
at Alice and send its second subsystem sequentially. Every
transmitted qubit has marginal population $m/s$, including after
earlier transmissions. The same estimate gives
\begin{equation}
 \ERinf(J_{s,m}^{\gamma,q})\leq s\,u_{\gamma,q}(m/s).
 \label{eq:gad-resource-upper}
\end{equation}
This is an upper estimate for the generalized resources. Unlike
Eqs.~\eqref{eq:REEexact} and \eqref{eq:resourceadditivity}, it
asserts neither equality nor additivity.

\section{Generalized amplitude-damping benchmark formulas}
\label{sec:gad-numerics}

We specify the established benchmark constructions used in
Fig.~\ref{fig:gad-bounds}. They concern the same GAD convention
and count rates per physical channel use. All entropies and
logarithms below are in bits.

\subsection{Channel output and the established hashing reference}

For $\varphi_x$ of Eq.~\eqref{eq:Omegax}, write
\begin{equation}
 \Omega_{\gamma,q,x}=(\id\otimes\GAD)(\varphi_x)
 =\begin{pmatrix}
 a&0&0&c\\0&b&0&0\\0&0&d&0\\c&0&0&e
 \end{pmatrix},
 \label{eq:gad-omega}
\end{equation}
in the ordered basis $00,01,10,11$, where
\begin{align}
 a&=(1-x)(1-\gamma q),~~b=(1-x)\gamma q,\nonumber\\
 d&=x\gamma(1-q),~~e=x[1-\gamma(1-q)],\nonumber\\
 c&=\sqrt{(1-\gamma)x(1-x)}.
 \label{eq:gad-omega-entries}
\end{align}
Its eigenvalues are $b,d$ and
\begin{equation}
 \mu_\pm=\frac{a+e\pm\sqrt{(a-e)^2+4c^2}}2.
 \label{eq:gad-omega-spectrum}
\end{equation}
Writing $S_\Omega=S(\Omega_{\gamma,q,x})$ and
$y=(1-\gamma)x+\gamma q$, the coherent and reverse coherent
information of this input are
\begin{equation}
 I_{\rm c}(x)=h_2(y)-S_\Omega,~~
 I_{\rm R}(x)=h_2(x)-S_\Omega.
 \label{eq:gad-hashing-informations}
\end{equation}
The displayed achievable reference is
\begin{equation}
 L_{\rm hash}(\gamma,q)=
 \max_{0\leq x\leq1}\max\{0,I_{\rm c}(x),I_{\rm R}(x)\}.
 \label{eq:gad-hashing-reference}
\end{equation}
This uses established hashing and reverse-hashing
protocols~\cite{DW,HOW,PGPBL,RCI}. It is a diagonal-input
benchmark, without an optimality claim for general distillation.
No new sector-distillation lower bound is used in this manuscript.

\subsection{Max-relative entropy and approximate covariance}

The analytic max-relative-entropy bound of Ref.~\cite{KSW},
Proposition~10, is
\begin{align}
 U_{\max}^{\rm GAD}(\gamma,q)
 &=\left[\log_2\!\left(1-\frac\gamma2+
                         \frac{D_{\gamma,q}}2\right)\right]_+,
 \nonumber\\
 D_{\gamma,q}&=\sqrt{\gamma^2(1-2q)^2+4(1-\gamma)}.
 \label{eq:gad-emax}
\end{align}
The positive part makes the bound zero precisely in the
entanglement-breaking region. It reduces to
$\log_2(2-\gamma)$ at $q=0,1$.

The approximate-covariance construction of Ref.~\cite{KSW},
Eq.~(208), compares the GAD channel to its unital member. With
\begin{align}
 \epsilon&=\gamma|q-1/2|,\nonumber\\
 g(t)&=(1+t)\log_2(1+t)-t\log_2t,
 \label{eq:gad-covariance-parameters}
\end{align}
the displayed curve is
\begin{equation}
 U_{\rm cov}^{\rm GAD}(\gamma,q)
 =\min\{1,P_\gamma+2\epsilon+g(\epsilon)\}.
 \label{eq:gad-approximate}
\end{equation}
We use continuous values at $t=0$. Individual benchmark curves
can remain positive in an entanglement-breaking region; the
independent zero-capacity criterion \eqref{eq:gad-eb} still applies.

\subsection{The two balanced-squashing constructions}

For a variable damping probability $t$, denote the ordinary and
upward channels by $\mathcal A_t^\downarrow$ and
$\mathcal A_t^\uparrow$. The two serial factorizations used in
Ref.~\cite{KSW}, Proposition~9, are
\begin{align}
 \GAD&=\mathcal A_{b_1}^\uparrow\circ
                     \mathcal A_{a_1}^\downarrow,\nonumber\\
 a_1&=\frac{\gamma(1-q)}{1-\gamma q},~~b_1=\gamma q,
 \label{eq:gad-serial-one}\\
 \GAD&=\mathcal A_{a_2}^\downarrow\circ
                     \mathcal A_{b_2}^\uparrow,\nonumber\\
 a_2&=\gamma(1-q),~~b_2=\frac{\gamma q}{1-\gamma(1-q)}.
 \label{eq:gad-serial-two}
\end{align}
These formulas are nonsingular in the interior; endpoint channels
are assigned their continuous limits.

For reproducibility, use the environmental basis fixed by
\begin{align}
 V_t^\downarrow\ket0&=\ket{00}_{BE},\nonumber\\
 V_t^\downarrow\ket1&=\sqrt{1-t}\ket{10}_{BE}
                         +\sqrt t\ket{01}_{BE},\nonumber\\
 V_t^\uparrow&=(X_B\otimes I_E)V_t^\downarrow X_A.
 \label{eq:gad-dilations}
\end{align}
Composing these dilations according to either serial order gives
an isometry $V_j:A\to BE_1E_2$, for $j=1,2$.
Let $\mathcal S=\mathcal A_{1/2}^\downarrow$ act separately on
each environment, with retained outputs $E'_1,E'_2$, and define
\begin{equation}
 \sigma_j(x)=
 (\id_B\otimes\mathcal S\otimes\mathcal S)
 \bigl[V_j((1-x)\proj0+x\proj1)V_j^\dagger\bigr].
 \label{eq:gad-squashed-state}
\end{equation}
Each balanced squashing isometry is symmetric under exchange of
its retained and discarded output. In a purification by $R$,
this symmetry gives
\begin{align}
 \tfrac12 I(R:B|E'_1E'_2)
 &=S(BE'_1E'_2)_{\sigma_j(x)}
       -S(E'_1E'_2)_{\sigma_j(x)}.
 \label{eq:gad-squashing-entropy}
\end{align}
Reference~\cite{KSW} proves that the input optimization for these
constructions can be restricted to the population $x$. Thus
\begin{equation}
 U_{{\rm sq},j}^{\rm GAD}
 =\max_{0\leq x\leq1}
 \bigl[S(BE'_1E'_2)_{\sigma_j(x)}-S(E'_1E'_2)_{\sigma_j(x)}\bigr].
 \label{eq:gad-squashing-bound}
\end{equation}
The plotted bound is
$\min\{U_{{\rm sq},1}^{\rm GAD},U_{{\rm sq},2}^{\rm GAD}\}$.
Each input maximization precedes the minimum over the two
constructions. No optimization over arbitrary squashing maps is
asserted.

The numerical procedure diagonalizes the density matrices in
Eqs.~\eqref{eq:gad-omega} and \eqref{eq:gad-squashed-state} and
performs scalar input optimizations, including the endpoints. As consistency checks, we verified numerically that both serial
dilations yield the state in Eq.~\eqref{eq:gad-omega}, that the convex decomposition reproduces the GAD channel, and that, at $q=0$, both squashing constructions recover the amplitude-damping bound in Eq.~\eqref{eq:sq}. At every sampled point, the upper bounds were at least as large as the displayed achievable rates. The relative ordering of the bounds is reported only for the sampled parameter values.

\end{document}